\documentclass[runningheads]{llncs}
\usepackage{times}
\usepackage{url}
\usepackage{amsmath,amssymb,latexsym}
\usepackage{algorithm}
\usepackage[noend]{algorithmic}
\algsetup{linenosize=\scriptsize}

\usepackage{paralist}
\usepackage{multicol}
\usepackage{hyperref}
\usepackage{wrapfig}
\usepackage{breakcites}
\usepackage{cancel}

\usepackage{color}
\usepackage[T1]{fontenc}
\usepackage{authblk}
\usepackage{soul}
\newcommand*{\affaddr}[1]{#1}
\newcommand*{\affmark}[1][*]{\textsuperscript{#1}}

\newtheorem{coro}[theorem]{Corollary}
\newtheorem{observation}[theorem]{Observation}

\renewcommand{\paragraph}[1]{\noindent{\bf #1}:\ }

\newenvironment{lemma-repeat}[1]{\begin{trivlist}
\item[\hspace{\labelsep}{\bf\noindent Lemma~\ref{#1} (repeated).}]\it}%
{\end{trivlist}}
\newenvironment{corollary-repeat}[1]{\begin{trivlist}
\item[\hspace{\labelsep}{\bf\noindent Corollary~\ref{#1} (repeated)}]\it}%
{\end{trivlist}}
\newenvironment{theorem-repeat}[1]{\begin{trivlist}
\item[\hspace{\labelsep}{\bf\noindent Theorem~\ref{#1} (repeated)}]\it}%
{\end{trivlist}}

\newcommand{\remove}[1]{}

\newcommand{\AlgName}{{\sc CCC}}

\newcommand{\setalglineno}[1]{%
	\setcounter{ALC@line}{\numexpr#1-1}}

\newcommand{\mathsc}[1]{{\normalfont\textsc{#1}}}

\newcommand{\blank}[1]{\hspace*{#1}}

\begin{document}

\title{Store-Collect in the Presence of Continuous Churn with
Application to Snapshots and Lattice Agreement\thanks{
    Supported by ISF grant 380/18 and NSF grant 1816922.}}
\author{Hagit Attiya\affmark[1],
        Sweta Kumari\affmark[1],
        Archit Somani\affmark[1],
        and Jennifer L. Welch\affmark[2]}
\institute{\affaddr{\affmark[1]Department of Computer Science, Technion, Israel}\\
        \email{\{hagit, sweta, archit\}@cs.technion.ac.il}\\
		\affaddr{\affmark[2]Department of Computer Science and Engineering, Texas A\&M University}\\
		\email{welch@cse.tamu.edu}}





\date{}
\authorrunning{Attiya, Kumari, Somani and Welch}
\titlerunning{Store-Collect in the Presence of Continuous Churn with Applications}

\maketitle

\begin{abstract}
We present an algorithm for implementing a store-collect object in an
asynchronous crash-prone message-passing dynamic system, where nodes
continually enter and leave.
The algorithm is very simple and efficient, requiring just one round trip
for a store operation and two for a collect.
We then show the versatility of the store-collect object for implementing
churn-tolerant versions of useful data structures, while shielding the
user from the complications of the underlying churn.
In particular, we present elegant and efficient implementations of 
atomic snapshot and generalized lattice agreement objects that 
use store-collect.


\end{abstract}

\section{Introduction}
\label{section:intro}

A popular programming technique that contributes to designing
provably-correct distributed applications is to use shared objects for
interprocess communication, instead of more low-level techniques.
Although shared objects are a convenient abstraction, they are not
generally provided in large-scale distributed systems; instead,
nodes keep copies of the data and communicate by
sending messages to keep the copies consistent.

\emph{Dynamic} distributed systems allow computing nodes to
enter and leave the system at will, either due to failures and recoveries,
moving in the real world, or changes to the systems' composition,
a process called \emph{churn}.
Motivating applications include those in
peer-to-peer, sensor, mobile, and social networks, as well as server farms.
We focus on the situation when the network is always fully
connected, which could be due to, say, an overlay network.
A broadcast mechanism is assumed through which a node can send a
message to all nodes present in the system;
the broadcast is not necessarily reliable and a message sent by a
failing node may not reach some of the nodes.

The usefulness of shared memory programming abstractions has been
long established for static systems
(e.g.,~\cite{Aspnes-notes,AttiyaBD1995}),
which have known bounds on the number of fixed computing nodes and
the number of possible failures.
This success has inspired work on providing the same for newer,
dynamic, systems.
However, most of this work has shown how to simulate a shared
read-write register (e.g.,
\cite{AguileraKMS2011,AttiyaCEKW2019,BaldoniBR2012,BaldoniBKR2009,GilbertLS2010}).
We discuss a couple of exceptions~\cite{BaldoniBR2016,KuznetsovRTP2019}
below.

In this paper, we promote the {\em store-collect} shared object
\cite{AttiyaFG2002} (defined in Section~\ref{section:sc problem})
as a primitive well-suited for
dynamic message-passing systems with an ever-changing set of participants.
Each node can store a value in a store-collect object with a
{\sc Store} operation and can collect the latest
value stored by each node with a {\sc Collect} operation.
Inherent in the specification of this object is an ability
to track the set of participants and to read their latest values.

Below we elaborate on three advantageous features of the store-collect
object:
The store-collect semantics is well-suited to dynamic systems
and can be implemented easily and efficiently in them;
the widely-used atomic snapshot object can be
implemented on top of a store-collect object; and
a variety of other commonly-used objects can be implemented either
directly on top of a store-collect or on top of an atomic snapshot object.
These implementations are simple and inherit the properties of
being churn-tolerant and efficient,
showing that store-collect combines algorithmic power and efficiency.

{\em A churn-tolerant store-collect object can be implemented
fairly easily.}
We adopt essentially the same system model as in \cite{AttiyaCEKW2019},
which allows ongoing churn as long as not too many churn events
take place during the length of time that a message is in transit.
To capture this constraint, there is an assumed upper bound $D$ on
the maximum message delay, but no (positive) lower bound.
Nodes do not know $D$ and have no local clocks, causing consensus to be
unsolvable \cite{AttiyaCEKW2019}.
The model differentiates between nodes that crash and nodes that leave;
nodes that have entered but not left are considered present even if crashed.
The number of nodes that can be crashed at any time
is bounded by a fraction of the number of nodes present at that time.
During any time interval of length $D$,
the number of nodes entering or leaving
is a fraction of the number of nodes present in
the system at the beginning of the interval.
(See Section~\ref{section:model} for model details.)

Our algorithm for implementing a churn-tolerant store-collect object is
based on the read-write register algorithm in \cite{AttiyaCEKW2019}.
It is simple and efficient: once a node joins,
it completes a store operation within one round-trip,
and a collect operation within two round-trips.
The store-collect object satisfies a variant of the ``regularity''
consistency condition, which is weaker than linearizability~\cite{HerlihyW1990}.
In contrast to our single-round-trip store operation, the write
operation in the algorithm of \cite{AttiyaCEKW2019} requires two round trips.
Another difference between the algorithms is that in ours, each
node keeps a local set of tuples with an entry for each known node
and its value instead of a single value; when receiving new information,
instead of overwriting the single value, our algorithm merges the new
information with the old.
One contribution of our work in this paper is a significantly revised
proof of the churn management protocol that is much simpler
than that in~\cite{AttiyaCEKW2019},
consequently making it easier to build on the results.
(See Section~\ref{section:algorithm}.)

{\em Building an atomic snapshot on top of a store-collect object is
easy!}  We present a simple algorithm with an elegant correctness proof
(Section~\ref{section:atomicSnap}).
One may be tempted to implement an atomic snapshot in our model
by plugging churn-tolerant registers (e.g., \cite{AttiyaCEKW2019})
into the original algorithm of \cite{AfekADGMG1993}.
Besides needlessly sequentializing accesses to the registers,
such an implementation would have to track the current set of participants.
A store-collect object which encapsulates the changing participants
and collects information from them in parallel,
yields a simple algorithm very similar in spirit to the original
but whose round complexity is linear instead of quadratic in
the number of participants.
The key subtlety of the algorithm is the mechanism for
detecting when a scan can be borrowed
in spite of difficulties caused by the
churn, in order to ensure termination.

Atomic snapshot objects have numerous uses in static systems,
e.g., to build multi-writer registers, concurrent timestamp systems,
counters, and accumulators,
and to solve approximate agreement and randomized consensus
(cf.~\cite{AfekADGMG1993,Aspnes-notes}).
In addition to analogous applications,
we show (Section~\ref{section:gla}) how
a churn-tolerant atomic snapshot object can be
used to provide a churn-tolerant generalized lattice agreement
object~\cite{FaleiroRRRV2012}.
This object supports a \textsc{Propose} operation whose argument is a
value belonging to a lattice and whose response is
a lattice value that is the join of some
subset of all prior input values, including its own argument.
Generalized lattice agreement is an extension of (single-shot)
\emph{lattice agreement},
well-studied in the static shared memory model~\cite{AttiyaHR1995}.
Generalized lattice agreement has been used to implement many
objects~\cite{ChordiaRRRV2013,FaleiroRRRV2012},
including atomic snapshots \cite{AttiyaHR1995},
and \emph{conflict-free replicated data
types}~\cite{KuznetsovRTP2019,ZhengGK2019,ZhengHG2018}, e.g.,
linearizable abort flags, sets, and max registers~\cite{KuznetsovRTP2019}.


{\em The store-collect object specification is versatile.}
Our atomic snapshot and generalized lattice agreement algorithms
demonstrate that layering linearizability on top of a
store-collect object is easy.
Yet not every application needs the costs associated with linearizability,
and store-collect gives the flexibility to avoid them.
Our approach to providing churn-tolerant shared objects is modular,
as the underlying complications of the message-passing and the churn is
hidden from higher layers by our store-collect implementation.
As evidence, we observe (Section~\ref{section:small apps}) that
store-collect allows very simple implementations of max-registers,
abort flags, and sets, in which an implemented
operation takes at most a couple of store and collect operations.
The choice of problems and the algorithms
follow \cite{KuznetsovRTP2019} but the algorithms inherit good
efficiency and churn-tolerance properties from our store-collect
implementation.

\paragraph{Related work}
An algorithm that directly implements an atomic snapshot object
in a {\em static message-passing system}, bypassing the use of registers,
is presented in~\cite{DelporteFRR2018}.
This algorithm includes several nice optimizations to improve the message
and round complexities.
These include speeding up the algorithm by parallelizing the
collect, as is already encapsulated in our store-collect algorithm.
Our atomic snapshot algorithm works in a \emph{dynamic} system and
has a shorter and simpler proof of linearizability.

Aguilera~\cite{Aguilera2004} presents a specification and algorithm
for atomic snapshots in a {\em dynamic model} in which nodes can continually
enter and communicate via {\em shared registers}.
This algorithm is then used for group membership and mutual exclusion in
that model.
Variations of the model were proposed in
\cite{GafniMT2001,MerrittT2000}, which provided algorithms for election,
mutual exclusion, consensus, collect, snapshot, and renaming.
%
Spiegelman and Keidar~\cite{SpiegelmanK2016} present atomic
snapshot algorithms
for a crash-prone dynamic system in which processes
communicate via shared registers.
%
Their algorithms
uniquely identify each scan operation with a version number to help
determine when a scan can be borrowed;
we use a similar mechanism in our snapshot algorithm.
However, our atomic snapshot algorithm uses a shared store-collect object
which tolerates \emph{ongoing} churn.
Our use of a non-linearizable building block requires a more delicate
approach to proving linearizability, as we cannot simply choose, say,
a specific write to an atomic register as the linearization point of
an update, as can be done in \cite{SpiegelmanK2016}.


The problem of implementing shared objects in the presence of
{\em ongoing} 
churn and crash failures in \emph{message-passing systems}
is studied in~\cite{BaldoniBKR2009,BaldoniBR2012},
which considers read-write registers,
and \cite{BaldoniBR2016},
which considers sets.
Unlike our results,
these papers assume the system size is restricted to a fixed window
and the system is eventually synchronous.
Like our algorithms, the set algorithm in \cite{BaldoniBR2016} uses
unbounded local memory at the nodes.


A popular alternative way to model churn
in \emph{message-passing systems} is as a sequence of quorum
configurations, each of which consists of a set of nodes and a quorum
system over that set (e.g.,
\cite{AguileraKMS2011,GafniM2015,GilbertLS2010,JehlVM2015,KuznetsovRTP2019}).
Explicit reconfiguration operations replace older configurations with
newer ones.
The assumptions made in
\cite{AguileraKMS2011,GafniM2015,GilbertLS2010,JehlVM2015,KuznetsovRTP2019}
are incomparable with those in \cite{AttiyaCEKW2019} and in our paper,
as the former assume churn eventually stops while the latter assume the
churn is bounded.


Most papers on generalized lattice agreement have assumed static systems
(cf.~\cite{AttiyaHR1995,ChordiaRRRV2013,FaleiroRRRV2012,ZhengGK2019,ZhengHG2018}.
A notable exception is~\cite{KuznetsovRTP2019}, which considers
dynamic systems subject to changes in the composition due to
reconfiguration.
This paper provides an implementation for a large class of shared
objects, including conflict-free replicated data types, that can be
modeled as a lattice.
By showing how to view the state of the system as a lattice as well,
the paper elegantly combines the treatment of the reconfiguration
and the operations on the object.
Unlike our work, the algorithms in~\cite{KuznetsovRTP2019} require that
changes to the system composition eventually cease in order to ensure
progress.

\section{The Store-Collect Problem}
\label{section:sc problem}

A shared {\em store-collect object}~\cite{AttiyaFG2002} supports concurrent
{\em store} and {\em collect} operations performed by some set of
clients.  Each operation has an invocation and response.  For a {\em store}
operation, the invocation is of the form {\sc Store}$_p(v)$, where $v$
is a value drawn from some set and $p$ indicates the invoking client,
and the response is of the form {\sc Ack}$_p$, indicating that the
operation has completed.  For a {\em collect} operation, the invocation is
of the form {\sc Collect}$_p$ and the response is of the form {\sc
  Return}$_p(V)$, where $V$ is a {\em view}, that is, a set of
client-value pairs without repetition of client ids.  We use the
notation $V(p)$ to indicate $v$ if $\langle p, v \rangle \in V$ and
$\bot$ if no pair in $V$ has $p$ as its first element.

Informally, the behavior required of a store-collect object is that
each {\em collect} operation should return a view containing the latest
value stored by each client.
We do not require the \emph{store} and \emph{collect}
operations to appear to occur instantaneously, that is, the object is
not necessarily linearizable.
Instead, we give a precise definition of the required behavior that is
along the lines of \emph{interval linearizability}~\cite{CastanedaRR2018}
or the specification of \emph{regular} registers~\cite{Lamport86}.

A sequence $\sigma$ of invocations and responses of {\em store} and
{\em collect} operations is a {\em schedule} if, for each client id $p$,
the restriction of $\sigma$ to invocations and responses by $p$
consists of alternating invocations and matching
responses, beginning with an invocation.
Each invocation and its matching following response (if present)
together make an operation.
If the response of operation $op$ comes before the invocation of operation
$op'$ in $\sigma$, then we say $op$ {\em precedes} $op'$ (in $\sigma)$
and $op'$ {\em follows} $op$.
We assume that every value written in a {\em store} operation in a
schedule is unique (a condition that can be achieved using sequence
numbers and client ids).


Given two views $V_1$ and $V_2$ returned by two {\em collect}
operations in a schedule $\sigma$, we write $V_1 \preceq V_2$ if, 
for every $\langle p, v_1 \rangle \in V_1$, 
there exists $v_2$ such that $\langle p, v_2 \rangle \in V_2$
such that either $v_1 = v_2$ or
the {\sc Store}$_p(v_1)$ invocation does not occur after 
the response of {\sc Store}$_p(v_2)$ in $\sigma$.

A schedule $\sigma$ satisfies {\em regularity for the store-collect problem}
if:
\begin{itemize}
\item For each {\em collect} operation {\it cop} in $\sigma$ that returns
$V$ and every client $p$, if $V(p) = \bot$, 
then no {\em store} operation by $p$ precedes {\it cop} in $\sigma$.
If $V(p) = v \ne \bot$, then
there is a {\sc Store}$_p(v)$ invocation that occurs in $\sigma$
before {\it cop} completes and no other {\em store} operation by $p$ occurs
in $\sigma$ between this invocation and the invocation of {\it cop}.

\item For every two {\em collect} operations in $\sigma$,
{\it cop}$_1$ which returns $V_1$ and {\it cop}$_2$ which returns $V_2$,
if {\it cop}$_1$ precedes {\it cop}$_2$ in $\sigma$,
then $V_1 \preceq V_2$.


\end{itemize}

\section{System Model}
\label{section:model}

We model each node $p$ as a state machine with a set of states,
containing two initial states $s_p^i$ and $s_p^{\ell}$.
Initial state $s_p^i$ is used if $p$ is initially in the system,
whereas $s_p^{\ell}$ is used if $p$ enters the system later.
The set of all nodes that are initially in the system  is denoted by $S_0$.
It is finite and nonempty.

State transitions are triggered by the occurrences of events.
Possible triggering events are:
entering the system ({\sc Enter}$_p$),
leaving the system ({\sc Leave}$_p$),
receipt of a message $m$ ({\sc Receive}$_p(m)$),
invocation of an operation ({\sc Collect}$_p$ or {\sc Store}$_p(v)$), and
crashing ({\sc Crash}$_p$).

A {\em step} of a node $p$ is a 5-tuple $(s',T,m,R,s)$ where $s'$ is the
old state, $T$ is the triggering event, $m$ is the message to
be sent, $R$ is a response ({\sc Return}$_p(V)$, {\sc Ack}$_p$, or
{\sc Joined$_p$}) or $\bot$, and $s$ is the new state.
The values of $m$, $R$ and $s$
are determined by a transition function applied to
$s'$ and $T$.
{\sc Return}$_p(V)$ is the response to {\sc Collect}$_p$,
{\sc Ack}$_p$ is the response to  {\sc Store}$_p$,
and
{\sc Joined}$_p$ is the response to {\sc Enter}$_p$.
If $T$ is {\sc Crash}$_p$,
then $m$ is $\bot$ and $R$ is $\bot$.

A {\em local execution} of a node $p$ is a sequence of steps such that:
\begin{itemize}
\item the old state of the first step is an initial state;
\item the new state of each step equals the old state of the next step;
\item if the old state of the first step is $s_p^i$, then no
      {\sc Enter}$_p$ event occurs;
\item if the old state of the first
      step is $s_p^{\ell}$, then
      the triggering event in the first step is {\sc Enter}$_p$ and
      there is no other occurrence of {\sc Enter}$_p$; and
\item at most one of {\sc Crash}$_p$ and {\sc Leave}$_p$ occurs and if so,
      it is in the last step.
\end{itemize}
In our model, a node that leaves the system cannot re-enter with the same id.
It can, however, re-enter with a new id.
Likewise, a node that crashes does not recover.
A node that crashes and recovers, but loses its state, can re-enter with a
new id.
Because nodes cannot measure time, a node that crashes and recovers,
retaining its state, can be treated as if no crash occurred.

A point in time is represented by a nonnegative real number.
A {\em timed local execution} is a local execution whose steps occur
at nondecreasing times.
If a local execution is infinite, the times at which its steps occur
must increase without bound.
Given a timed local execution of a node, if $(s',T,m,R,s)$ is the step with the
largest time less than or equal to $t$,
then $s$ is the {\em state} of that node {\em at time $t$}.
A node $p$ is said to be {\em present} at time $t$ if it entered the
system (i.e., its first step has time at most $t$)
but has not left (i.e., {\sc Leave}$_p$ does not occur at or before $t$).
The number of nodes that are present at time $t$ is denoted by $N(t)$.
A {\em crashed} node (i.e., a node for which {\sc Crash}$_p$ occurs at
or before $t$) is still considered to be present.
A node is said to be {\em active} at time $t$
if it is present and not crashed at $t$.

An {\em execution} $e$ is a possibly infinite set of timed local
executions,
one for each node that is ever present in the system,
such that there is a nonempty finite set of nodes that
are initially members.
Formally, the first step of each node $p \in S_0$ occurs at time 0 and
the first step of each other node occurs after time 0.

We assume a reliable broadcast communication service that
provides nodes with a mechanism to send the same message
to all nodes\footnote{
    Sending a message to a single recipient can be accomplished
    by broadcasting the message and indicating the intended
    recipient so that others will ignore the message.}
in the system; message delivery is FIFO.
However, if a broadcast invocation is the last thing that a node
does before crashing, the message is not guaranteed to be received
by all the nodes; this is a weaker assumption than that made
in~\cite{AttiyaCEKW2019}.
If a message $m$ sent at time $t$ is received by a node at time $t'$,
then the {\em delay} of this message is $t'-t$.
This encompasses transmission delay as well as time for handling
the message at both the sender and receiver.
Let $D > 0$ denote the {\em maximum message delay} that can occur in the
system.
Formally:
\begin{itemize}
\item
Every sent message has at most one matching receipt at
each node and every message receipt has exactly one matching message send.

\item
If a message $m$ is sent by a node $p$ at time $t$,
$p$'s next event is not {\sc Crash}$_p$,
and node $q$ is active throughout $[t,t+D]$ (i.e., $q$ enters by time $t$
and does not leave or crash by time $t+D$), then
$q$ receives $m$.
The delay of every received message is in $(0,D]$.

\item
Messages from the same sender are received in the order they are sent
(i.e., if node $p$ sends message $m_1$ before sending message $m_2$,
then no node receives $m_2$ before it receives $m_1$).
This can be achieved by tagging each message with
the id of its sender and a sequence number.
\end{itemize}

Let $\alpha > 0$ and $0 < \Delta \leq 1$ be real numbers that denote the
{\em churn rate} and {\em failure fraction}, respectively.
Let $N_{min}$ be a positive integer, the minimum system size.
The parameters $\alpha$ and $\Delta$ are known to the nodes,
but $N_{min}$ and $D$ are not.
We assume executions satisfy three assumptions:
\begin{description}
\item[Churn Assumption]
For all times $t>0$,
there are at most $\alpha \cdot N(t)$ {\sc Enter} and {\sc Leave}
events in $[t,t+D]$.

\item[Minimum System Size]
For all times $t \ge 0$, $N(t) \ge N_{min}$.

\item[Failure Fraction Assumption]
For all times $t \ge 0$,
at most $\Delta \cdot N(t)$ nodes are crashed at time $t$.

\end{description}

A node $p$ is said to be a {\em member} at time $t$ if it has {\em joined} the
system (i.e., $p \in S_0$ or {\sc Joined}$_p$ occurs at or before $t$)
but has not left (i.e., {\sc Leave}$_p$ does not occur at or before $t$).
Note that, at any time $t$, the members are a subset of the present nodes.
It is possible that some members have crashed.

We assume ``well-formed'' interactions between client threads and
their users:
An invocation occurs at node $p$
only if $p$ has already joined but has not left or crashed,
i.e., $p$ is a member.
Furthermore, no previous invocation at $p$ is pending,
i.e., at most one operation is pending at each node.

An algorithm is a {\em correct implementation of a store-collect
object} in our model if the following are true for all
executions with well-formed interactions:
\begin{itemize}
\item For every node $p \notin S_0$, if
{\sc Enter}$_p$ occurs, then at least one of {\sc Leave}$_p$,
{\sc Crash}$_p$, or {\sc Joined}$_p$ occurs subsequently; that is, every
node that enters the system and remains active eventually joins.
For every node $p \in S_0$, {\sc Joined}$_p$ never occurs.

\item For every node $p$, if {\sc Store}$_p(v)$ (respectively,
{\sc Collect}$_p$) occurs, then
at least one of {\sc Leave}$_p$, {\sc Crash}$_p$, or {\sc Ack}$_p$
(respectively, {\sc Return}$_p(V)$)
occurs subsequently; that is, every store or collect operation
invoked at a node that remains active eventually completes.

\item The schedule resulting from the restriction of the execution to
the {\em store} and {\em collect} invocations and responses
satisfies regularity for the store-collect problem.
\end{itemize}

\section{The Continuous Churn Collect (\AlgName{}) Algorithm}
\label{section:algorithm}

In our algorithm, nodes run \emph{client} threads, which invoke
\emph{collect} and \emph{store} operations, and \emph{server} threads.
We assume that the code segment that is executed in response to each
event executes without interruption.

To track the composition of the system
(Algorithm~\ref{algo:Common}),
a node $p$ maintains a set {\em Changes} of
events concerning the nodes that have entered the system.
When an {\sc Enter}$_p$ event occurs,
$p$ adds \emph{enter(p)} to its \emph{Changes} set (Line \ref{line:present1})
and broadcasts an \textbf{enter} message requesting
information about prior events (Line \ref{line:present2}).
When $p$ finds out that another node $q$ has entered the system, either by
receiving an \textbf{enter} message directly from $q$ or by receiving
an \textbf{enter-echo} message for $q$ from a third node, it
adds \emph{enter(q)} to its \emph{Changes} set
(Line \ref{line:present3} or \ref{line:present6}).
When $p$ receives an \textbf{enter} message from a node $q$,
it replies with an \textbf{enter-echo} message containing
its \emph{Changes} set,
its current estimate {\em LView (local view)} of the state of the simulated object,
its flag {\em is\_joined} indicating whether $p$ has joined yet,
and the id $q$ (Line \ref{line:present4}).
The first time that $p$ receives an \textbf{enter-echo}
in response to its own \textbf{enter} message (i.e., one that ends with
its own id) from a joined node, it computes \emph{join\_threshold},
the number of
\textbf{enter-echo} messages it needs to get before it can join (Line
\ref{line:present9}) and increments its \emph{join\_counter}
(Line \ref{line:present10}).

The fraction $\gamma$ is used to calculate \emph{join\_threshold},
the number of \textbf{enter-echo} messages that should be received
before joining,
based on the size of the \emph{Present} set (nodes that have entered,
but have not left, see Line \ref{line:present9}).
Setting $\gamma$ is a key challenge in the algorithm as setting it too
small might not propagate updated information, whereas setting it too
large might not guarantee termination of the join.

\remove{
	When $q$ has received sufficiently many messages in reply to its
	request, it knows relatively accurate information about prior
	events.
	The fraction $\gamma$ is used to calculate the number of messages that
	should be received before joining (stored in the local variable \emph{join\_threshold}),
	based on the size of the \emph{Present} set (nodes that have entered, but have not left, see Line \ref{line:present9}).
	Setting $\gamma$ is a key challenge in the algorithm as setting it too
	small might not propagate updated information, whereas setting it too
	large might not guarantee termination of the join.
	
	When $q$ receives an enter-echo in reply (i.e., that ends with $q$),
	it (*decrements its {\em join\_threshold}*) (Line \ref{line:present10}).
	The first time $q$ receives such an enter-echo from a joined node,
	it computes \emph{join\_threshold},
	the number of enter-echo messages it needs to get before it can join (Line \ref{line:present9}).
}

When the required number of replies to the \textbf{enter} message sent
by $p$ is received (Line \ref{line:present11}), $p$ adds
\emph{join(q)} to its \emph{Changes} set, sets its \emph{is\_joined}
flag to \emph{true} (Line \ref{line:present12}), broadcasts a message
saying that it has joined (Line \ref{line:present14}) and outputs
{\sc Joined}$_p$ (Line \ref{line:present15}).  When $p$ finds out that
another node $q$ has joined, either by receiving a \textbf{join} message
directly from $q$
or by receiving a \textbf{join-echo} message for $q$ from a third node, it
adds \emph{join(q)} to its \emph{Changes} set (Line \ref{line:present16}
or \ref{line:present19}).  When a {\sc Leave}$_p$ event occurs, $p$
broadcasts a \textbf{leave} message (Line \ref{line:present21}) and
halts (Line \ref{line:present22}). When $p$ finds out that another node $q$ is
leaving the system, either by receiving a \textbf{leave} message
directly from $q$ or by receiving a \textbf{leave-echo} message for $q$
from a third node, it adds \emph{leave(q)} to its \emph{Changes} set (Line
\ref{line:present23} or \ref{line:present25}).

Initially, node $p$'s \emph{Changes} set equals
$\{\it{enter}(q) | q \in S_0\} \cup \{\it{join}(q) | q \in S_0\}$,
if $p \in S_0$,
and $\emptyset$ otherwise.  Node $p$ also maintains a set of nodes that it
believes are present: \emph{Present} $= \{q | \it{enter}(q) \in
\it{Changes} \wedge \it{leave}(q) \not\in \it{Changes}\}$,
i.e.,~nodes that have entered, but have not
left, as far as $p$ knows. Essentially, Algorithm \ref{algo:Common} of
\AlgName{} is the same as {\sc CCReg} \cite{AttiyaCEKW2019} except for
Line \ref{line:present5},
which merges newly received information with current local information
instead of overwriting it.

Once a node has joined, its client thread can handle \emph{collect}
and \emph{store} operations (Algorithm \ref{algo:cs_client}) and
its server thread (Algorithm \ref{algo:Server}) can respond to clients.
The client at node $p$ maintains a derived variable \textit{Members} =
$\{q ~|~$ \emph{join(q)} $\in$ \emph{Changes}$ \wedge$
\emph{leave(q)} $\not\in$ \emph{Changes}$\}$ of nodes that $p$
considers as members, i.e., nodes that have joined but not left.

Our implementation adds a sequence number, \emph{sqno}, to each value
in a view, which is now a set of triples, $\{\langle p, v,
sqno\rangle, \ldots \}$, without repetition of node ids.
We use the notation $V(p) = v$ if there exists {\it sqno} such that
$\langle p, v, {\it sqno} \rangle \in V$, and $\perp$ if no triple
in $V$ has $p$ as its first element.

A merge of two views, $V_1$ and $V_2$, picks the latest value stored
by each node according to the highest \emph{sqno}.
If a triple for
node $p$ is in $V_2$ but not in $V_1$ then the merge includes the
triple for $p$ from $V_2$ as well, and vice versa.
That is,

\begin{definition}
Given two views $V_1$ and $V_2$, {\it merge}$(V_1,V_2)$ is defined as the
subset of $V_1 \cup V_2$ consisting of every triple whose node id is in
one of $V_1$ and $V_2$ but not the other, and, for node ids that appear
in both $V_1$ and $V_2$, it contains only the triple with the larger
sequence number.
\end{definition}

Note that $V_1, V_2 \preceq {\it merge}(V_1,V_2)$.

\begin{algorithm}[tb]
	\begin{algorithmic}[1]
		\setalglineno{1}
		\small
		\item[] {\bf Local Variables:}
		\item[] {\it LView}:  set of (node id, value, sequence number)
		triples, initially $\emptyset$
		\COMMENT{local view}
		\item[] \emph{is\_joined}:  Boolean, initially false
		\COMMENT{true iff $p$ has joined the system}
		\item[] \emph{join\_threshold}:  int, initially 0
		\COMMENT{number of enter-echo messages needed for joining}
		\item[] \emph{join\_counter}:  int, initially 0
		\COMMENT{number of enter-echo messages received so far}
		\item[] \emph{Changes}:  set of {\it enter}$(q)$, {\it leave}$(q)$,
		and {\it join}$(q)$
		\COMMENT{active membership events known to $p$}
		\item[] \hspace*{1cm} initially $\{${\it enter}$(q) | q \in S_0\}$ $\cup$
		$\{${\it join}$(q) | q \in S_0\}$ if $p \in S_0$,
		and $\emptyset$ otherwise
		\item[] {\bf Derived Variable:}
		\item[] $\mbox{\it Present} = \{q ~|~$ \emph{enter(q)} $\in
		\mbox{\it \emph{Changes}} \wedge$ \emph{leave(q)} $\not\in
		\mbox{\it Changes} \}$
		\item[] \hrulefill	
		\setlength{\multicolsep}{0pt}	
		\begin{multicols*}{2}
			\item[] \blank{-.45cm}\bf When {\sc Enter}$_p$ occurs:
			\STATE add $\mbox{\it enter(p)}$ to $\mbox{\it Changes}$ \label{line:present1}
			\STATE broadcast $\langle$enter, $p\rangle$ \label{line:present2}
			\item[]	
			\item[] \blank{-.55cm} \bf When {\sc Receive}$_p \langle$enter, $q\rangle$ occurs:
			\STATE add $\mbox{\it enter(q)}$ to $\mbox{\it Changes}$ \label{line:present3}
			\STATE broadcast$\langle$enter-echo,$\mbox{\it Changes}$, $\mbox{\it LView}$, \\$\mbox{\it is\_joined}$, $\mbox{\it q}\rangle$	 \label{line:present4}
			\item[]	
			
			\item[] \blank{-.55cm} \bf When {\sc Receive}$_p \langle$enter-echo, $\mbox{\it C}$, $\mbox{\it RView}$, $\mbox{\it j}$, $\mbox{\it q}\rangle$ \\\blank{-.55cm} occurs:
			\STATE $\mbox{\it LView}$ = $\mbox{\it merge(LView, RView)}$ \label{line:present5}
			\STATE $\mbox{\it Changes} = \mbox{\it Changes} \cup C$ \label{line:present6}
			\IF{$\neg \mbox{\it is\_joined} \wedge (p == q)$} \label{line:present7}
			\IF{$(j==$\mbox{\it true}$)\!\wedge\!(\mbox{\it join\_threshold}==0)$\\} \label{line:present8}
			\STATE 	$\mbox{\it join\_threshold} =  \gamma \cdot | \mbox{\it Present} | $ \label{line:present9}
			\ENDIF
			\STATE $\mbox{\it join\_counter}${\tt ++}  \label{line:present10}
			\IF{$ \mbox{\it join\_counter $\geq$ join\_threshold} > 0$\\} \label{line:check if enough enter echoes} \label{line:present11}
			\STATE $\mbox{\it is\_joined} = \mbox{\it true}$ \label{line:present12}
			\STATE add \mbox{\it join(p)} to $\mbox{\it Changes}$ \label{line:present13}
			\STATE broadcast $\langle$join, $p\rangle$ \label{line:present14}
			\STATE return {\sc Joined}$_p$ \label{line:present15}
			\ENDIF
			\ENDIF
			\item[]
			\item[] \blank{-.7cm} \bf When {\sc Receive}$_p \langle$join, $q\rangle$ occurs:
			\STATE add \mbox{\it join(q)} to $\mbox{\it Changes}$ \label{line:present16}
			\STATE add \mbox{\it enter(q)} to $\mbox{\it Changes}$ \label{line:present17}
			\STATE broadcast $\langle$join-echo, $q\rangle$ \label{line:present18}
			\item[]	
			\item[] \blank{-.7cm} \bf When {\sc Receive}$_p \langle$join-echo, $q\rangle$ occurs:
			\STATE add \mbox{\it join(q)} to $\mbox{\it Changes}$ \label{line:present19}
			\STATE add \mbox{\it enter(q)} to $\mbox{\it Changes}$ \label{line:present20}
			\item[]	
			\item[] \blank{-.7cm} \bf When {\sc Leave}$_p$ occurs:
			\STATE broadcast $\langle$leave, $p\rangle$ \label{line:present21}
			\STATE halt \label{line:present22}
			\item[]	
			
			\item[] \blank{-.7cm} \bf When {\sc Receive}$_p \langle$leave, $q\rangle$ occurs:
			\STATE add \mbox{\it leave(q)} to $\mbox{\it Changes}$ \label{line:present23}
			\STATE broadcast $\langle$leave-echo, $q\rangle$ \label{line:present24}	
			\item[]
			
			\item[] \blank{-.7cm} \bf When {\sc Receive}$_p \langle$leave-echo, $q\rangle$ occurs:
			\STATE add \mbox{\it leave(q)} to $\mbox{\it Changes}$	\label{line:present25}
		\end{multicols*}
	\end{algorithmic}
	\caption{\AlgName---Common code managing churn, for node $p$.}
	\label{algo:Common}
\end{algorithm}

Each node keeps a local copy of the current view in its {\em LView}
variable.
In a \emph{collect operation},
a client thread requests the latest value of servers' local views
using a \textbf{collect-query} message (Line~\ref{line:client4}).
When a server node $p$ receives a \textbf{collect-query}
message, it responds with its local view ({\em LView}) through a
\textbf{collect-reply} message (Line~\ref{line:server6}) if $p$ has
joined the system.  When the client receives a \textbf{collect-reply}
message, it merges its \emph{LView} with the \emph{received view (RView)},
to get the latest value corresponding to each node (Line~\ref{line:client6}).
Then the client waits for sufficiently many
\textbf{collect-reply} messages before broadcasting the current value of its
\emph{LView} variable in a \textbf{store} message (Line~\ref{line:client9}).
When server $p$ receives a \textbf{store} message with
a view \emph{RView}, it merges \emph{RView} with its local
\emph{LView} (Line~\ref{line:server1}) and, if $p$ is joined, it broadcasts
\textbf{store-ack} (Line~\ref{line:server3}).
The client waits for sufficiently many \textbf{store-ack} messages
before returning \emph{LView} to complete the {\em collect}
(Line~\ref{line:client21});
this threshold is recalculated in Line~\ref{line:recalc} to reflect
possible changes to the system composition that the client has observed.

\begin{algorithm}[tb]
	\begin{algorithmic}[1]
		\setalglineno{26}
		\small
		\item[] {\bf Local Variables:}
		\item[] \emph{optype}: string, initially $\bot$
		\COMMENT{indicates which type of operation (\emph{collect}
			or \emph{store}) is pending}
		\item[] \emph{tag}:  int, initially 0
		\COMMENT{counter to identify currently pending operation by $p$}
		\item[] \emph{threshold}:  int, initially 0
		\COMMENT{number of replies/acks needed for current phase}
		\item[] \emph{counter}:  int, initially 0
		\COMMENT{number of replies/acks received so far for current phase}
		\item[] \emph{sqno}:  int, initially 0
		\COMMENT{sequence number for values stored by $p$}
		\item[] {\bf Derived Variable:}
		\item[] \emph{Members} $= \{q |$ \emph{join(q)} $\in$
		\emph{Changes} $\wedge$ \emph{leave}$(q) \not\in$
		\emph{Changes}$\}$
		\item[] \hrulefill	
		\setlength{\multicolsep}{0pt}	
		\begin{multicols*}{2}
			\small	
			\item[] \blank{-.7cm} \bf When {\sc Collect}$_p$ occurs:
			\STATE \mbox{\it optype} $=$ \mbox{\it collect;} \mbox{\it tag}++ \label{line:client1}
			\STATE $\mbox{\it threshold} = \beta \cdot |\mbox{\it Members}| $ \label{line:client2}
			\STATE \mbox{\it counter $=$ 0}
			\STATE broadcast $\langle$collect-query, \mbox{\it tag}, $p\rangle$  \label{line:client4}
			\item[]	
			\item[] \blank{-.7cm} \bf When {\sc Receive}$_p \langle$collect-reply, $\mbox{\it RView, t, q} \rangle$ \\\blank{-.6cm}occurs:
			\IF{\mbox{\it (t == tag)} $\wedge (q == p)$} \label{line:client5}
			\STATE $\mbox{\it LView}$ = $\mbox{\it merge(LView, RView)}$ \label{line:client6}
			\STATE $\mbox{\it counter}${\tt ++} \label{line:client7}
			\IF{(\mbox{\it counter $\geq$ threshold})}   \label{line:client8}
			\STATE ${\it threshold} = \beta \cdot |{\it Members}|
			$
			\label{line:recalc}
			\STATE \mbox{\it counter $=$ 0}
			\STATE broadcast $\langle$store, $\mbox{\it LView, tag, p}\rangle$ \label{line:client9}
			\item[]
			\ENDIF
			\ENDIF
			\columnbreak
			\item[] \blank{-.7cm} \bf When {\sc Store}$_p(v)$ occurs:
			\STATE \mbox{\it optype} $=$ \mbox{\it store;} \mbox{\it tag}++ \label{line:client11}
			\STATE $\mbox{\it sqno}$++ \label{line:client12}
			\STATE $\mbox{\it LView} = \mbox{\it merge(LView,} \{\langle p, v,
			\mbox{\it sqno} \rangle \})$ \label{line:client13}
			\STATE $\mbox{\it threshold} = \beta \cdot |\mbox{\it Members}|
			$ \label{line:client14}
			\STATE \mbox{\it counter $=$ 0}
			\STATE broadcast $\langle$store, $\mbox{\it LView, tag, p} \rangle$ \label{line:client16}
			\item[]
			
			\item[] \blank{-.7cm} \bf When {\sc Receive}$_p \langle$store-ack, \mbox{\it t}, $q \rangle$ occurs:
			\IF{\mbox{\it (t $==$ tag)} $\wedge (q==p)$} \label{line:client17}
			\STATE $\mbox{\it counter}${\tt ++}  \label{line:client18}
			\IF{($\mbox{\it counter $\geq$ threshold}$)} \label{line:client19}
			\STATE if (\mbox{\it optype$==$store}) then return \mbox{\sc ACK} \label{line:client20}
			\STATE else return \mbox{\it LView} \label{line:client21} 	
			\ENDIF
			\ENDIF
		\end{multicols*}
	\end{algorithmic}
	\caption{\AlgName---Client code, for node $p$.}
	\label{algo:cs_client}
\end{algorithm}

In a \emph{store operation}, a client thread
updates its local variable \emph{LView} to reflect
the new value by doing a merge (Line~\ref{line:client13}) and
broadcasts a \textbf{store} message (Line~\ref{line:client16}).
When server $p$ receives a \textbf{store} message with view \emph{RView},
it merges \emph{RView} with its local \emph{LView} (Line~\ref{line:server1})
and, if $p$ is joined, it broadcasts \textbf{store-ack} (Line~\ref{line:server3}).
The client waits for sufficiently many \textbf{store-ack} messages before
completing the {\em store} (Line~\ref{line:client20}).

\begin{algorithm}
\begin{algorithmic}[1]
	\setalglineno{48}
	\setlength{\multicolsep}{0pt}
\begin{multicols*}{2}
		\small
	\item[] \blank{-.7cm} \bf When {\sc Receive}$_p \langle$store, $\mbox{\it RView,tag},q\rangle$ occurs:\\
	\STATE $\mbox{\it LView}$ $=$ $\mbox{\it merge(LView, RView)}$ \label{line:server1}
	\IF{\mbox{\it is\_joined}} \label{line:server2}
	        \STATE
	        broadcast
	         $\langle$store-ack, \mbox{\it tag}, $q \rangle$ \label{line:server3}
	\ENDIF
	\STATE broadcast $\langle$store-echo, $\mbox{\it LView} \rangle$ \label{line:server4}
	\item[]
	\item[] \blank{-.7cm} \bf When {\sc Receive}$_p \langle$collect-query, \mbox{\it tag}, $q\rangle$ occurs:
	\IF{\mbox{\it is\_joined}} \label{line:server5}
		\STATE
		broadcast
		 $\langle$collect-reply, $\mbox{\it LView, tag}, q \rangle$ \label{line:server6}
	\ENDIF
	\item[]
	\item[] \blank{-.7cm} \bf When {\sc Receive}$_p \langle$store-echo, $\mbox{\it RView} \rangle$ occurs: \\
	\STATE $\mbox{\it LView}$ $=$ $\mbox{\it merge(LView, RView)}$ \label{line:server7}
\end{multicols*}
\end{algorithmic}
\caption{\AlgName---Server code, for node $p$.}
\label{algo:Server}
\end{algorithm}

The fraction $\beta$ is used to calculate the number of messages that
should be received (stored in local variable \emph{threshold}) based
on the size of the \emph{Members} set, for the operation to terminate.
Setting $\beta$ is a key challenge in the algorithm as setting it
too small might not return correct information from \emph{collect} or
\emph{store}, whereas setting it too large might not guarantee
termination of the \emph{collect} and \emph{store}.

We define a {\em phase} to be the execution
by a client node $p$ of one of the following intervals of its code:
\begin{itemize}
\item lines \ref{line:client1} through \ref{line:client8},
the first part of a {\em collect} operation,
\item lines \ref{line:recalc} through \ref{line:client9}
and \ref{line:client17} through \ref{line:client21},
the second part of a \emph{collect} operation called
the ``store-back'', or
\item lines \ref{line:client11} through \ref{line:client20},
the entirety of a \emph{store} operation.
\end{itemize}
The first kind of phase is called a \emph{collect phase}
while the second and third kinds are called a \emph{store phase}.

For any completed phase $\varphi$ executed by node $p$,
define {\it view}$(\varphi)$ to be the value of {\it LView}$_p^t$,
where $t$ is the time at the end of the phase.
Since a {\em store} operation consists solely of a store phase,
we also apply the notation to an entire {\em store} operation.





\newcommand{\zed}{Z}

\section{Proof of \AlgName{} Store-Collect Algorithm}
\label{section:sc-proof}

To prove the correctness of the algorithm,
consider any execution of the algorithm.
The correctness of the algorithm relies on the following constraints
($Z = \left[ (1 - \alpha)^3\right.$ $\left.- \Delta \cdot (1 + \alpha)^3 \right]$,
which is  
the fraction of nodes that survive an interval of length $3D$):
\begin{align}
N_{min} &\ge \frac{1}{Z + \gamma - (1 + \alpha)^3}
   \label{parameters:C} \displaybreak[0]\\
\gamma  &\le Z/(1+\alpha)^3
   \label{parameters:D} \displaybreak[0]\\
\beta   &\le Z/(1 + \alpha)^2
   \label{parameters:E} \displaybreak[0]\\
\beta   &> \frac{(1-Z)(1+\alpha)^5 + (1+\alpha)^6}
              {((1-\alpha)^3- \Delta \cdot (1+\alpha)^2) ((1+\alpha)^2+1)}
   \label{parameters:H}
\end{align}
Fortunately, there are values for the parameters $\alpha$, $\Delta$,
$\gamma$, and $\beta$ that satisfy these constraints.
In the extreme case when $\alpha = 0$ (i.e., no churn), the failure
fraction $\Delta$ can be as large as 0.21; in this case, it suffices
to set both $\gamma$ and $\beta$ to 0.79 for any value of $N_{min}$ that
is at least 2.
As $\alpha$ increases up to 0.04, $\Delta$ must decrease approximately
linearly until reaching 0.01;
in this case, it suffices to set $\gamma$ to 0.77 and $\beta$ to 0.80
for any value of $N_{min}$ that is at least 2.
%
The following technical claims hold: 

\begin{lemma}
\label{lem:changing-size}
For all $i \in \mathbb{N}$ and all $t \ge 0$, \\
(a) at most $((1+\alpha)^i-1)\cdot N(t)$ nodes enter during
$(t,t+i\cdot D]$; and \\
(b) $N(t+i\cdot D) \le (1 + \alpha)^i \cdot N(t)$.
\end{lemma}

\begin{proof}
	The proof is by induction on $i$.
	
	{\em Basis:} $i = 0$.  For all $t$, the interval $(t,t+0\cdot D] =
	(t,t]$ is empty and (a) and (b) are true.
	
	{\em Induction:}  Assume (a) and (b) are true for $i$ and show
	for $i+1$.
	Partition the interval $(t,t+(i+1) \cdot D]$ into $(t, t+D]$ and
	$(t+D, t + (i+1) \cdot D]$.
	Since the latter interval is of length $i \cdot D$, the inductive
	hypothesis applies (replacing $t$ with $t+D$) and we get:\\
	(a) at most $((1+\alpha)^i - 1) \cdot N(t+D)$ nodes enter during
	$(t+D,t+ (i+1) \cdot D]$; and \\
	(b) $N(t+(i+1) \cdot D) \le (1 + \alpha)^i \cdot N(t+D)$.
	
	By the churn assumption, (i) at most $\alpha \cdot N(t)$ nodes enter during $(t,t+ D]$ and thus (ii) $N(t+ D) \le (1+\alpha) \cdot N(t)$.
	To show (a) for $i+1$,
	combine (i) with the inductive hypothesis for part (a) to see that
	the number of nodes that enter during $(t,t+(i+1) \cdot D]$ is
	\begin{align*}
		&\le \alpha \cdot N(t) + ((1+\alpha)^i - 1) \cdot N(t+D)  \\
		&\le \alpha \cdot N(t) + ((1+\alpha)^i - 1) \cdot (1+\alpha) \cdot N(t)
		\qquad \mbox{by (ii)}\\
		&= \alpha \cdot N(t) + (1+\alpha)^{i+1} \cdot N(t) - (1+\alpha) \cdot N(t) \\
		&= ((1+\alpha)^{i+1}-1) \cdot N(t).
	\end{align*}
	
	To show (b) for $i+1$:
	\begin{align*}
		N(t+(i+1) \cdot D) &\le (1+\alpha)^i \cdot N(t+D) \qquad \mbox{by the inductive
			hypothesis for (b)} \\
		&\le (1+\alpha)^i \cdot (1+\alpha) \cdot N(t) \qquad \mbox{by (ii)} \\
		&= (1+\alpha)^{i+1} \cdot N(t).
	\end{align*}
	\qed\end{proof}

Calculating the maximum number of nodes that can leave in an interval
of length $i \cdot D$ as a function of the number of nodes at the
beginning of the interval (i.e., the analog of part (a) of Lemma
\ref{lem:changing-size}) is somewhat complicated by the possibility
of nodes entering during the interval, allowing additional nodes to leave.

\begin{lemma}
	\label{lem:max-leave}
	For all $\alpha$, $0 < \alpha < .206$,
	all non-negative integers $i \le 3$, and every time $t \ge 0$, at most
	$(1 - (1 - \alpha)^i) \cdot N(t)$ nodes leave during $(t, t+ i \cdot D]$.
\end{lemma}

\begin{proof}
	The proof is by induction on $i$.
	
	{\em Basis:}  $i = 0$.  For all $t$, the interval $(t, t + 0 \cdot D] =
	(t,t]$ is empty and so no nodes leave during it.
	
	{\em Induction:}  Suppose the lemma is true for $i$ and prove it for $i+1$.
	Partition the interval $(t,t+(i+1) \cdot D]$ into $(t, t+D]$ and
	$(t+D, t + (i+1) \cdot D]$.
	Since the latter interval is of length $i \cdot D$, the inductive
	hypothesis applies (replacing $t$ with $t+D$), stating that
	the number of nodes that leave in the latter interval is at most
	$(1 - (1 - \alpha)^i) \cdot N(t+D)$.
	
	Let $e$ be the {\em exact} number of nodes that enter in $(t,t+D]$
	and $\ell$ be the {\em exact} number of nodes that leave in $(t,t+D]$.
	The number of nodes that leave in the entire interval is:
	\begin{align*}
		&\le \ell + (1 - (1 - \alpha)^i) \cdot N(t+D) \qquad \mbox{by the inductive
			hypothesis} \\
		&\le \ell + (1 - (1 - \alpha)^i) \cdot \left[ (1 + \alpha) \cdot N(t) - 2 \ell \right]
	\end{align*}
	The last line is true since $N(t+D) = N(t) + e - \ell$ which equals
	$N(t) + (\ell + e) - 2 \ell$, which is
	at most $N(t) + \alpha \cdot N(t) - 2 \ell$ by the churn assumption.
	Algebraic manipulations show that this is
	\begin{align*}
		&\le (1 - (1 - \alpha)^i) \cdot (1 + \alpha) \cdot N(t) + (2 (1-\alpha)^i - 1)
		\ell \\
		&\le (1 - (1 - \alpha)^i) \cdot (1 + \alpha) \cdot N(t) +
		(2(1-\alpha)^i - 1) \cdot \alpha \cdot N(t)
	\end{align*}
	The last line is true since $\ell \le \alpha \cdot N(t)$ by the churn
	assumption and
	$(2(1-\alpha)^i - 1)$ is non-negative by the constraints on $\alpha$ and $i$
	in the premise of the lemma.
	This expression equals $(1 - (1 - \alpha)^{i+1}) \cdot N(t)$.\qed
\end{proof}

Recall that a node is {\em active} at time $t$ if it has entered, but
not left or crashed, by time $t$.
The next lemma counts how many of the nodes that are active at a given
time are still active after $3D$ time has elapsed.
It introduces the quantity $\zed$ which is the fraction of nodes that
must survive an interval of length $3D$.

\begin{lemma}
\label{lem:active-3D}
For any interval $[t_1,t_2]$ with $t_2 - t_1 \le 3D$,
where $S$ is the set of nodes present at $t_1$, at least
$\zed \cdot |S|$ of the nodes in $S$ are active at $t_2$.
(Recall that
$\zed =\left[ (1 - \alpha)^3 - \Delta \cdot (1 + \alpha)^3 \right]$.)
\end{lemma}

\begin{proof}
	Consider any interval $[t_1,t_2]$ with $t_2 - t_1 \le 3D$ and let $S$
	be the set of nodes present at $t_1$.
	
	By Lemma~\ref{lem:max-leave}, at most $(1-(1-\alpha)^3) \cdot |S|$ nodes
	leave during the interval.
	In the worst case, all of the leavers are among the original set of nodes
	$S$.
	
	By Lemma~\ref{lem:changing-size}, part (b), the number of nodes present at
	$t_2$ is at most $(1 + \alpha)^3 \cdot |S|$.
	By the crash assumption, up to a $\Delta$ fraction of them crash,
	and in the worst case all of these are among the original
	set of nodes $S$.
	
	Thus the number of nodes in $S$ that remain active at the end of the
	interval is at least
	\begin{align*}
		|S| - (1 - (1 - \alpha)^3)\cdot |S| - \Delta \cdot (1 + \alpha)^3 \cdot |S|
		= \left[(1-\alpha)^3 - \Delta \cdot (1 + \alpha)^3\right] \cdot |S|.
	\end{align*}
	\qed \end{proof}

As an immediate corollary, since $|S|$ must be at least $N_{min}$,
the lower bound on $N_{min}$ given in Constraint~\ref{parameters:C}
shows that at least one node survives.
To match its use cases, the corollary is stated with respect to a
time that is in the middle of the interval.

\begin{coro}
	\label{cor:at-least-one-active}
	For every $t > 0$, at least one node is active throughout the interval
	$[\max\{0,t-2D\},t+D]$.
\end{coro}

Throughout the proof,
a local variable name is subscripted with $p$ and
superscripted with $t$ to denote its value in node $p$
at time $t$; e.g., $v_p^t$ is the value of node $p$'s
local variable $v$ at time $t$.

In the analysis, we will frequently be comparing the data in nodes'
{\it Changes} sets to the set of {\sc Enter}, {\sc Joined}, and {\sc Leave}
events that have actually occurred in a certain interval.
We refer to these as {\em membership events}.
We are especially interested in these events that trigger
a broadcast invoked by a node that is not in the middle of crashing,
as these broadcasts are guaranteed to be received by all nodes that
are present for the requisite interval.
We call these {\em active membership events.}
Because of the assumed initialization of the nodes in $S_0$, we
use the convention that the set of active
membership events occurring in the interval $[0,0]$ is
$\{ {\it enter}(p) | p \in S_0\} \cup \{ {\it join}(p) | p \in S_0\}$.

The next lemmas describe how a node's {\it Changes} set relates to prior
active membership events.
Lemma~\ref{lem:present-changes} states that a node that has been present in the
system sufficiently long (at least $2D$ time), has all the
information about active membership events up until $D$ time in the past.
Lemma~\ref{lem:joined-changes} states that a joined node, no matter how
recently it entered the system, has all the
information about active membership events up until $2D$
time in the past.  The later parts of the correctness proof only
use Lemma~\ref{lem:joined-changes}, but its proof relies on
Lemma~\ref{lem:present-changes}.
The proof of Lemma~\ref{lem:joined-changes} relies on
Lemma~\ref{lem:enter-echo-from-long-lived},
which is rather technical and states that under certain circumstances
a node receives an enter-echo message from a long-lived node;
we have extracted it as a separate lemma
as it is also used later in the proof of Lemma~\ref{lem:joined-stores}.
Throughout the proof we denote by $t_p^e$ the time when event
{\sc Enter}$_p$ occurs.

The proofs of Lemmas~\ref{lem:present-changes} and \ref{lem:joined-changes}
use the next observation, which
follows from the fact that nodes broadcast enter/join/leave messages
when they enter/join/leave and these messages take at most $D$ time to
arrive at active nodes (unless the broadcast is the very last step
by a crashing node, in which case the message might not be received by
some nodes).

\begin{observation}
	\label{obs:5}
	For every node $p$ and all times $t \ge t_p^e + D$ such that $p$ is active
	at time $t$, {\it Changes}$_p^t$ contains all the active
	membership events for $[t_p^e,t-D]$.
\end{observation}

\begin{lemma}
	\label{lem:present-changes}
	For every node $p$ and all times $t \ge t_p^e + 2D$
	such that $p$ is active at $t$,
	{\it Changes}$_p^t$ contains all the active membership events for $[0,t-D]$.
\end{lemma}

\begin{proof}
	The proof is by induction on the order in which nodes enter.
	In particular, we consider the nodes in increasing order of {\sc Enter}
	events, breaking ties arbitrarily, and show the properties
	are true for the current node at all relevant times.
	
	{\em Basis:}  The first nodes to enter are those in $S_0$
	and they are assumed to do so at time 0.
	Consider $p \in S_0$.
	For $t \ge 2D$, Observation~\ref{obs:5} gives the result.
	
	{\em Induction:} Let $p$ be the next node (not in $S_0$) to enter,
	at time $t_p^e$, and assume the lemma is true
	for all nodes that entered previously.
	
	Consider any time $t \ge t_p^e + 2D$ such that $p$ is active at $t$.
	By Corollary~\ref{cor:at-least-one-active}, there exists a node $q$ that is
	active throughout $[t_p^e - 2D, t_p^e + D]$.
	Let $t'$ be the time when $q$ receives $p$'s enter message and
	$t''$ be the time when $p$ receives $q$'s enter-echo response.
	We will show that {\it Changes}$_p^t$ contains all the active membership
	events for $[0,t-D]$ in three steps: one for $[0,t'-D]$, one for
	$[t'-D,t_p^e]$, and one for $[t_p^e,t-D]$.
	
	\begin{enumerate}
		
		\item Note that $q$ enters the system at least $2D$ time before
		it sends its enter-echo message to $p$ at time $t'$.  By the
		inductive hypothesis, when $q$ sends that message, its
		{\it Changes} set contains all the active
		membership events for $[0,t'-D]$.
		Once $p$ receives
		the message, at time $t''$ which is less than or equal to $t$, its
		{\it Changes} set also contains all the active
		membership events for $[0,t'-D]$.
		
		\item
		Suppose some node $r$ enters, joins, or leaves in $[t'-D,t_p^e]$
		and $r$ does not crash during that event.
		Node $r$'s enter/join/leave message is received by $q$ either
		before $t_p^e$, in which case the information is included in $q$'s
		enter-echo message to $p$,
		or after $t_p^e$, in which case $q$ sends an enter/join/leave-echo
		message for $r$, which is received by $p$ before $t$.
		In either case, the information about $r$'s event propagates to $p$
		before $t$.
		Thus the result holds for $[t'-D,t_p^e]$.
		
		\item Observation~\ref{obs:5} gives the result for $[t_p^e,t-D]$. \qed
	\end{enumerate}
\end{proof}

\begin{lemma}
	\label{lem:enter-echo-from-long-lived}
	Suppose node $p$ is joined and active at some time $t$ and the first
	enter-echo response that $p$ receives from a joined node $q$ is
	sent at time $t' \le t$.
	If {\it Changes}$_q^{t'}$ contains all the active membership events for
	$[0,\max\{0,t'-2D\}]$,
	then before $p$ joins, it receives an enter-echo response from some
	node $q'$ that is active throughout the interval $[\max\{0,t'-2D\},t'+D]$.
\end{lemma}

\begin{proof}
	Let $S$ be the set of nodes present at time $\max\{0,t'-2D\}$.
	We will show that at least one of the enter-echo responses received by
	$p$ before joining is from a node in $S$, which is our desired $q'$.
	We start with the value of {\it join\_threshold}, which is the number of
	enter-echo responses for which $p$ waits before joining, and then subtract
	(1) the maximum
	number of enter-echo responses that could come from nodes not in $S$,
	(2) the maximum number of nodes in $S$ that could leave too soon
	(before $t'+D$), and
	(3) the maximum number of nodes in $S$ that could crash too soon
	(before $t'+D$).
	
	The value of {\it join\_threshold} is based on the size
	of $p$'s {\em Present} set at time $t''$, immediately after $p$ receives
	the enter-echo response from $q$
	(cf.\ Line \ref{line:present9} of Algorithm 1).
	By the premise of the lemma,
	{\it Changes}$_q^{t'}$ contains all the active membership events for
	$[0,\max\{0,t'-2D\}]$.
	Thus when $p$ receives the enter-echo response from $q$ at time $t'' \le
	t'+D$, its {\em Present}
	variable contains, at a minimum, all the nodes in $S$ minus
	those that left during $[\max\{0,t'-2D\},t'']$---call this
	quantity $\ell$---and minus those that crashed while broadcasting their
	enter message so that $p$ did not receive it---call this quantity $f$.
	Thus {\it join\_threshold} $\ge \gamma \cdot (|S| - \ell - f)$.
	
	We now count the maximum number of enter-echo responses that $p$
	can receive from nodes not in $S$ before joining.
	These would be from
	nodes that enter after $\max\{0,t'-2D\}$ but no later than $t'+D$, as nodes
	entering after $t'+D$ do not receive $p$'s enter message.
	The number of such nodes
	is $((1+\alpha)^3 - 1) \cdot |S|$ by Lemma~\ref{lem:changing-size}
	part (a).
	
	We then subtract the maximum number of nodes in $S$ that leave before
	$t'+D$.
	By Lemma~\ref{lem:max-leave}, at most
	$(1 - (1 - \alpha)^3) \cdot |S|$ nodes leave during
	$(\max\{0,t'-2D\},t'+D]$.
	In the worst case, all these leavers are in $S$.
	Recall that we have already charged $\ell$ to this budget.
	
	Finally we subtract the maximum number of nodes in $S$ that crash
	before $t'+D$.
	The system size at $t'+D$ is at most $(1+\alpha)^3 \cdot |S|$
	by Lemma~\ref{lem:changing-size} part (b).
	At most $\Delta \cdot (1 + \alpha)^3 \cdot |S|$ nodes are crashed
	at time $t'+D$ by the crash assumption.
	In the worst case, all these crashed nodes are in $S$.
	Recall that we have already charged $f$ crashes to this budget.
	
	What remains is at least
	\begin{align*}
		\gamma \cdot (|S| - \ell - f)
		&- ((1+\alpha)^3 - 1) \cdot |S| \\
		& - [(1 - (1-\alpha)^3) \cdot |S| - \ell]
		- [\Delta \cdot (1 + \alpha)^3 \cdot |S| - f]
	\end{align*}
	which after doing some algebra and using fact that $(1 - \gamma)(\ell+f)
	\ge 0$ is equal to
	\begin{align*}
		|S| \cdot (\gamma - (1 + \alpha)^3 + (1 - \alpha)^3 -
		\Delta(1 + \alpha)^3)
	\end{align*}
	
	Since $|S|$ must be be at least $N_{min}$, Constraint~\ref{parameters:C}
	ensures that the expression is at least one.
	Thus before $p$ joins, it receives an enter-echo response from at least
	one node $q'$ that is active throughout
	$(\max\{0,t'-2D\},t'+D]$.
	\qed
\end{proof}

\begin{lemma}
\label{lem:joined-changes}
For every node $p$ and all times $t$ such that $p$ is joined and active
at $t$,
{\it Changes}$_p^t$ contains all the active membership events for
$[0,\max\{0,t-2D\}]$.
\end{lemma}

\begin{proof}
	The proof is by induction on the order in which nodes join.
	In particular, we consider the nodes in increasing order of {\sc Join}
	events, breaking ties arbitrarily, and show the properties
	are true for the current node at all relevant times.
	
	{\em Basis:}  The first nodes to join are those in $S_0$
	and they are assumed to do so at time 0.
	Consider $p \in S_0$.
	When $t \le 2D$, we just need to show that {\it Changes}$_p^t$
	contains all the active membership events for $[0,0]$,
	which is true by the assumed initialization of nodes in $S_0$.
	When $t > 2D$, Observation~\ref{obs:5} implies the result.
	
	{\em Induction:}  Let $p$ be the next node (not in $S_0$) to join
	and assume the lemma is true for all nodes that previously joined.
	Consider any time $t$ when $p$ is joined and active.
	
	When $t - t_p^e \ge 2D$, Lemma~\ref{lem:present-changes} gives the result.
	So we suppose $t - t_p^e < 2D$.
	If $t \le 2D$, then all that's required is for {\it Changes}$_p^t$
	to include all the active membership events in $[0,0]$.
	Since $p$ joined, it received an enter-echo message from some previously
	joined node, which by the inductive hypothesis had all the active
	membership
	events for $[0,0]$ in its {\it Changes} set when it sent the enter-echo.
	Thus $p$ receives all the active
	membership events for $[0,0]$ before it joins.
	For rest of the proof, assume $t > 2D$.
	
	We will show that {\it Changes}$_p^t$ contains all the active
	membership events
	for $[0,t-2D]$  in two steps:  one for $[0,\max\{0,t'-2D\}]$ and one for
	$[\max\{0,t'-2D\},t-2D]$ for an appropriately chosen $t' < t$.
	
	\begin{enumerate}
		\item Let $q$ be the first joined node from which $p$ gets an enter-echo
		response to its enter message.
		Let $t'$ be the time when $q$ sends the enter-echo message.
		By the inductive hypothesis, since $q$ is joined at $t'$,
		{\it Changes}$_q^{t'}$ contains all the active membership events for
		$[0,\max\{0,t'-2D\}]$, and thus so does {\it Changes}$_p^t$.
		
		\item By Lemma~\ref{lem:enter-echo-from-long-lived},
		$p$ receives an enter-echo message at some
		time before it joins from a node $q'$ that is active throughout the
		interval $[\max\{0,t'-2D\},t'+D]$.
		Let $u'$ be the time when $q'$ sends its enter-echo response to $p$.
		Suppose some node $r$ enters, joins or leaves in $[\max\{0,t'-2D\},
		t-2D]$ and does not crash during the event.
		Our goal is to show that $p$ receives the information about
		$r$ by time $t$.
		The latest that $r$'s message is sent is $t-2D$.
		Since $t-t' \le t - t_p^e < 2D$, it
		follows that $t-D \le t' + D$ and thus $q'$ is guaranteed to receive
		$r$'s message, as $q'$ is still active at $t-D$, the latest that the
		message could arrive.
		If $q'$ receives $r$'s message before $u'$,
		then $p$ gets the information about $r$ by time $t$ via the enter-echo
		response from $q'$.
		Otherwise, $q'$ receives $r$'s message after $u'$; the latest this can
		be is $t-D$.
		Then $q'$ sends an enter-echo message for $r$ which is received by $p$
		by time $t$. \qed
	\end{enumerate}
\end{proof}

We can prove that a node that is active sufficiently long eventually joins.

\begin{theorem}
\label{thm:join-liveness}
Every node $p$ that enters at some time $t$ and is active for
at least $2D$ time joins by time $t+ 2D$.
\end{theorem}

\begin{proof}
	The proof is by induction on the order in which nodes enter the system.
	
	{\em Basis:}  The first nodes to enter are those in $S_0$
	and they are assumed to do so at time 0.
	Since they also are assumed to join at time 0, the theorem follows.
	
	{\em Induction:} Let $p$ be the next node (not in $S_0$) to enter,
	at time $t_p^e$, and assume the lemma is true
	for all nodes that entered previously.
	Suppose $p$ is active at $t_p^e + 2D$.
	
	First we show that $p$ receives an enter-echo response to its enter
	message from at least one joined node.
	
	Suppose $t_p^e < 2D$.
	By Corollary~\ref{cor:at-least-one-active},
	at least one node in $S_0$ is active throughout
	$[0,3D]$ and thus responds to $p$'s enter message.
	
	Suppose $t_p^e \ge 2D$.
	By Corollary~\ref{cor:at-least-one-active},
	there is a node $q$ that is active throughout $[t_p^e - 2D, t_p^e + D]$.
	Then $q$ enters at least $2D$ time before $t_p^e$ and by the
	inductive hypothesis is joined by $t_p^e$.
	Since $q$ is active at least until $t_p^e + D$, it receives $p$'s enter
	message by time $t_p^e + D$ and sends back an enter-echo which is received
	by $p$ by time $t_p^e + 2D$.
	
	We now calculate an upper bound on
	{\it join\_threshold}, the number of
	enter-echo responses for which $p$ waits before joining.
	This value is based on the size of $p$'s {\it Present} set when it
	first receives an enter-echo response from a joined node (cf.\ Line
	\ref{line:present9} of Algorithm 1).
	Let $q'$ be the sender of this message, let $t'$ be the time when
	the message is sent and $t''$ the time when it is received.
	Since $t' \ge t_p^e \ge 2D$, it follows that $t'-2D \ge 0$.
	By Lemma~\ref{lem:joined-changes}, {\it Changes}$_{q'}^{t'}$ contains
	all the active membership events for $[0,t'-2D]$
	and thus so does {\it Changes}$_p^{t''}$.
	As a result, {\it Present}$_p^{t''}$ contains, at most,
	all the nodes that are present at time $t'-2D$
	(call this set $S$) plus the maximum set of nodes that could have
	entered since then.
	Since $t'' \le t' + D$, it follows from Lemma~\ref{lem:changing-size}
	part (a)
	that at most $((1+\alpha)^3 - 1) \cdot |S|$ nodes enter during
	$(t'-2D,t'']$.
	Thus {\it join\_threshold} $\le \gamma \cdot (1 + \alpha)^3 \cdot |S|$.
	
	We now show that $p$ is guaranteed to receive at least {\it join\_threshold}
	enter-echo responses from nodes in $S$ by time $t_p^e + 2D$.
	Each node in $S$ that does not leave or crash by $t_p^e + D$
	receives $p$'s enter message
	and sends an enter-echo response by time $t_p^e+D$, which is received
	by $p$ by time $t_p^e+2D$.
	The minimum number of such nodes is, by Lemma~\ref{lem:active-3D}
	and considering the interval $(t'-2D,t'+D]$:
	\begin{align*}
	\zed \cdot |S|
	&\ge \gamma \cdot (1 + \alpha)^3 \cdot |S| \qquad \mbox{by Constraint~\ref{parameters:D}} \\
	&\ge {\it join\_threshold}.
	\end{align*}
\qed
\end{proof}

We prove that a phase terminates if the invoking client node is active long enough.

\begin{theorem}
\label{thm:phase-liveness}
A phase invoked by a client that remains active completes within $2D$ time.
\end{theorem}

\begin{proof}
Consider a phase invoked by node $p$ at time $t$.
We show that the number of nodes that respond to $p$'s collect-query or store
message is at least as large as the value of {\it threshold} computed by $p$
in Line~\ref{line:client2} or~\ref{line:recalc} or~\ref{line:client14} of
Algorithm 2.

Let $S$ be the set of nodes present at time $\max\{0,t-2D\} = t'$.
By Lemma~\ref{lem:active-3D}, the number of those nodes that are still
active at time $t+D$ is at least $Z \cdot |S|$.
If $t' = 0$, then $S = S_0$ and all these nodes are joined throughout;
otherwise, by Theorem~\ref{thm:join-liveness} all these nodes are joined
by time $t$.

We now show that $|S| \ge |Present_p^t|/(1 + \alpha)^2$.
By Lemma~\ref{lem:joined-changes}, {\it Changes}$_p^t$ contains
all the active membership events for $[0,t']$.
{\it Present}$_p^t$ is as large as possible if
every node in $S$ succeeds in the broadcast of its enter message,
none of the nodes
in $S$ leave during $[t',t]$, and the maximum number of nodes
enter during that interval and their enter messages get to $p$ by time $t$.
Lemma~\ref{lem:changing-size} part (a) implies that the maximum number of nodes
that can enter is $(1 + \alpha)^2 \cdot |S|$.
Thus $|Present_p^t| \le (1 + \alpha)^2 \cdot |S|$.

Thus the number of nodes that are
joined by time $t$ and are still active at time $t+D$,
guaranteed to respond to $p$, is at least
\begin{align*}
\zed \cdot |S|
  &\ge \zed \cdot |{\it Present}_p^t| / (1 + \alpha)^2 \\
  &\ge \beta \cdot |{\it Present}_p^t|
                        \qquad \mbox{by Constraint~\ref{parameters:E}} \\
  &\ge \beta \cdot |{\it Members}_p^t|
\end{align*}
The last inequality holds since {\it enter}$(q)$ is added 
to {\it Changes}$_p$ together with {\it join}$(q)$.
Since {\it threshold} is set to $\beta \cdot |${\it Members}$_p^t|$
at time $t$, $p$ receives the
required number of collect-reply or store-ack messages by time $t+2D$
and the phase completes.\qed
\end{proof}

The following observation is true since in this case node $p$ receives
phase $s$'s store message directly within $D$ time.

\begin{observation}
	\label{obs:direct-ts-info}
	For any store phase $s$ that starts at time $t_s$ and
	calls broadcast (Line~\ref{line:client16} of Algorithm~\ref{algo:cs_client})
	without crashing during the broadcast,
	and any node $p$
	that is active throughout $[t_s, t]$ where $t \ge t_s + D$,
	{\it view}$(s) \preceq$ {\it LView}$_p^t$.
\end{observation}

The next lemma is the analog of Lemma~\ref{lem:present-changes}: a node
that has been active for at least $2D$ time ``knows about'' store
phases that started up to $D$ in the past.

\begin{lemma}
	\label{lem:present-stores}
	If node $p$ is active at any time $t \ge t_p^e + 2D$, then
	{\it view}$(s) \preceq$ {\it LView}$_p^t$ for every
	store phase $s$ that starts at or before $t-D$ and
	calls broadcast (Line~\ref{line:client16} of Algorithm~\ref{algo:cs_client})
	without crashing during the broadcast.
\end{lemma}

\begin{proof}
	The proof is by induction on the order in which nodes enter the system.
	
	{\em Basis:}  The first nodes to enter are those in $S_0$ and they do so
	at time 0.
	The claim holds by Observation~\ref{obs:direct-ts-info}.
	
	{\em Induction:} Let $p$ be the next node (not in $S_0$) to enter
	and assume the claim is true for all nodes that entered previously.
	Consider any time $t \ge t_p^e + 2D$ when $p$ is active.
	Let $s$ be a store phase that starts at $t_s \le t - D$
	and calls broadcast without crashing.
	If $t_s \ge t_p^e$, the claim holds by Observation~\ref{obs:direct-ts-info}.
	
	Suppose $t_s < t_p^e$.
	By Corollary~\ref{cor:at-least-one-active}, there is at least one node $q$
	that is active throughout $[\max\{0,t_p^e - 2D\}, t_p^e + D]$.
	Since $t \ge t_p^e + 2D$, $p$ receives $q$'s enter-echo response by time
	$t$.  Since views and sequence numbers are included in enter-echo messages,
	{\it LView}$_q^{t'} \preceq$ {\it LView}$_p^t$,
	where $t'$ is the time when $q$ receives $p$'s enter message.
	
	{\em Case 1:}  $t_s < \max\{0,t_p^e - D\}$.
	We show that the inductive hypothesis applies for node $q$, time $t'$,
	and store phase $s$.  Thus
	{\it view}$(s) \preceq$ {\it LView}$_q^{t'}$,
	and by transitivity,
	{\it view}$(s) \preceq$ {\it LView}$_p^t$.
	To show that the inductive hypothesis holds, note that $q$ enters
	before $p$, $q$ has been active for at least $2D$ time by $t'$ and
	store phase $s$ starts at or before $t'-D$.
	
	{\em Case 2:}  $t_s \ge \max\{0,t_p^e - D\}$.
	The store message sent during $s$ is guaranteed to arrive at $q$ either
	before $t_p^e$ or at or after $t_p^e$.
	In the former case, $q$'s enter-echo response, which $p$ receives
	by $t_p^e + 2D \le t$, contains a view $V$ such that {\it view}$(s)
	\preceq V$.
	In the latter case, $q$'s store-echo message contains a view $V$ with
	{\it view}$(s) \preceq V$ and $p$ receives this message
	by $t_s + 2D < t_p^e + 2D \le t$.
	In both situations, {\it view}$(s) \preceq$ {\it LView}$_p^t$.\qed
\end{proof}

The next lemma is the analog of Lemma~\ref{lem:joined-changes}:  a node that
is joined ``knows about'' store phases that started up to $2D$ in the past.

\begin{lemma}
	\label{lem:joined-stores}
	If node $p$ is joined and active at any time $t$, then
	{\it view}$(s) \preceq$ {\it LView}$_p^t$ for every store phase $s$
	that starts at or before $t-2D$ and
	calls broadcast (Line~\ref{line:client16} of Algorithm~\ref{algo:cs_client})
	without crashing.
\end{lemma}

\begin{proof}
	The proof is by induction on the order in which nodes join the system.
	
	{\em Basis:}  The first nodes to join are those in $S_0$
	and they do so at time 0, which is also the time that they enter.
	The claim holds by Observation~\ref{obs:direct-ts-info}.
	
	{\em Induction:} Let $p$ be the next node (not in $S_0$) to join
	and assume the claim is true for all nodes that joined
	previously.
	Consider any time $t$ at which $p$ is joined and active.
	Let $s$ be any store phase that starts at $t_s \le t - 2D$
	and calls broadcast without crashing.
	If $t \ge t_p^e + 2D$, then the claim follows from
	Lemma~\ref{lem:present-stores}.
	
	Suppose $t < t_p^e + 2D$.
	For every store phase that starts at or after $t_p^e$,
	the claim follows from Observation~\ref{obs:direct-ts-info}.
	
	Consider any store phase that starts at some time $t_s < t_p^e$.
	Let $q$ be the sender of the first enter-echo response received by $p$
	from a joined node; suppose the message is sent at $t'$ and received at $t''$.
	
	{\em Case 1:}  $t_s < t' - 2D$.
	We show that the inductive hypothesis holds for node $q$, time $t'$,
	and store phase $s$.  Thus
	{\it view}$(s) \preceq$ {\it LView}$_q^{t'}$, and by transitivity,
	{\it view}$(s) \preceq$ {\it LView}$_p^t$.
	To show that the inductive hypothesis holds, note that $q$ joins before $p$,
	it is joined at time $t'$, and store phase $s$ starts before $t'-2D$.
	
	{\em Case 2:}  $t_s \ge t' - 2D$.
	Since $q$ is joined at $t'$, Lemma~\ref{lem:joined-changes} implies that
	{\it Changes}$_q^{t'}$ contains all the active membership events for
	$[0,\max\{0,t'-2D\}]$.
	Thus Lemma~\ref{lem:enter-echo-from-long-lived} applies and
	before $p$ joins it receives an enter-echo
	response from a node $q'$ that is active throughout $[\max\{0,t'-2D\},
	t'+D]$.
	The store message sent during $s$ is guaranteed to arrive at $q'$ either
	before $t_p^e$ or at or after $t_p^e$. In the former case, the enter-echo message from $q'$ that is sent to $p$
	contains a view $V$ with {\it view}$(s) \preceq V$; this message
	is received by $p$ before it joins.
	In the latter case, the store-echo message from $q'$ that is sent
	to $p$ contains a view $V$ with {\it view}$(s) \preceq V$; this message
	is received by $p$ by $t_s + 2D < t_p^e + 2D \le t$.
	In both situations, {\it view}$(s) \preceq$ {\it LView}$_p^t$.\qed
\end{proof}

The next lemma gives a lower bound on the size of a node's {\it Members}
set as a function of the size of the system $3D$ time in the past.

\begin{lemma}
	\label{lem:members-lb}
	For every node $p$ and every time $t$ at which $p$ is joined and active, \\
	$|${\it Members}$_p^t| \ge ((1 - \alpha)^3 - \Delta \cdot (1 + \alpha)^2)
	\cdot N(\max\{0,t-3D\})$.
\end{lemma}

\begin{proof}
	Let $S$ be the set of nodes that are present at time $\max\{0,t-3D\}$,
	so $|S| = N(\max\{0,t-3D\})$.
	
	First, assume $t \ge 3D$. 
	Since Theorem~\ref{thm:join-liveness} implies that it takes at most $2D$
	time for a node to join,
	at a minimum {\it Members}$_p^t$ contains all the nodes in $S$ except
	for those that leave during $[t-3D,t]$ and those that crash
	while broadcasting their join message so that $p$ does not receive the
	message.
	By Lemma~\ref{lem:max-leave}, the maximum number of nodes that leave
	during $[t-3D,t]$ is $(1 - (1 - \alpha)^3) \cdot |S|$.
	To maximize the number of nodes that crash while sending their join
	message, we consider the largest that the system can be by time
	$t - D$ (the latest that the nodes in $S$ can join), which is
	$(1 + \alpha)^2 \cdot |S|$ by Lemma~\ref{lem:changing-size}.
	We assume that
	the maximum number of nodes crash, which is
	$\Delta \cdot (1 + \alpha)^2 \cdot |S|$,
	and that all the crashed nodes are in $S$.
	Thus,
	$|${\it Members}$_p^t| \ge ((1 - \alpha)^3 - \Delta \cdot (1 + \alpha)^2)
	\cdot |S|$.
	
	Now, assume $t < 3D$. 
	Then $S$ equals $S_0$, the set of nodes initially
	in the system, and {\it Members}$_p^t$ is minimized if no nodes enter and
	the maximum number of nodes in $S_0$ leave and $p$ receives all
	their leave messages by time $t$.
	Since Lemma~\ref{lem:max-leave} implies that the maximum number of nodes that
	can leave during $[0,t]$ is $(1 - (1 - \alpha)^3 ) \cdot |S|$,
	it follows that
	$|${\it Members}$_p^t| \ge (1 - \alpha)^3 \cdot |S|$, which is
	bigger than the desired lower bound.
	%
	\qed
\end{proof}


To prove the following lemmas, we consider two cases:
If the two phases are sufficiently far apart
in time, then an information-propagation argument, analogous to that
used for the {\it Changes} sets, applies.
If the two phases are close together in time, then an argument
relating to overlapping sets of contacted nodes is used.


\begin{lemma}
\label{lem:views-ordered-by-time}
For any store phase $s$ and any collect phase $c$, if $s$ finishes
before $c$ starts and $c$ terminates,
then {\it view}$(s) \preceq$ {\it view}$(c)$.
\end{lemma}

\begin{proof}
	Let $p_1$ be the client node that executes $s$ and
	$t_s$ the start time of $s$.
	Let $p_2$ be the client node that invokes $c$ and
	$t_c$ the start time of $c$.
	Let $Q_s$ be the set of nodes that $p_1$ hears from during $s$
	(i.e., that sent messages causing $p_1$ to increment {\it counter}
	on Line \ref{line:client18} of Algorithm 2) and
	$Q_c$ be the set of nodes that $p_2$ hears from during $c$
	(i.e., that sent messages causing $p_2$ to increment {\it counter}
	on Line \ref{line:client7} of Algorithm 2).
	
	{\em Case I:}  $t_c - t_s \ge 2D$.
	Consider any node $q \in Q_c$.
	Since $q$ is in $Q_c$, $q$ is joined when it receives $c$'s collect-query
	message at some time, say $t \ge t_c$.
	By the assumption of the case, $t - t_s \ge 2D$.
	Thus by Lemma~\ref{lem:joined-stores},
	{\it view}$(s) \preceq$ {\it LView}$_q^t$.
	Since $p_2$ receives an enter-echo message from $q$ containing
	{\it LView}$_q^t$ before completing $c$, it follows that
	{\it view}$(s) \preceq$ {\it view}$(c)$.
	
	{\em Case II:} $t_c - t_s < 2D$.
	We will show that $Q_c$ and $Q_s$ have a nonempty intersection
	and thus $Q_c$ contains a node whose {\it LView} variable is $\preceq$
	{\it view}$(s)$ before it sends its collect-reply message to $p_2$, ensuring
	that {\it view}$(s) \preceq$ {\it view}$(c)$.
	We define the following sets of nodes.
	\begin{itemize}
		\item Let $J$ be the set of all nodes that are joined and active at
		some time in $[t_c,t_c+D]$.
		These are the nodes that could possibly respond to $c$'s collect-query
		message.
		Thus $Q_c \subseteq J$.
		\item Let $K \subseteq Q_s$ be the set of nodes in $Q_s$ that are still active
		at $t_c$.  Note that $K \subseteq J$.
	\end{itemize}
	We will show that $|Q_c| + |K| > |J|$. Since $Q_c$ and $K$ are both
	subsets of $J$, it follows that they intersect, and thus $Q_c$ and $Q_s$
	intersect.
	We show the inequality by calculating an upper bound on $|J|$ and lower
	bounds on $|Q_c|$ and $|K|$.  All three bounds are stated in
	terms of a common quantity, which is the system size at a particular time
	$t^* = \max\{0,t_c - 2D\}$.
	
	First we calculate an upper bound on $|J|$.
	Since it takes at most $2D$ time to join after entering by
	Theorem~\ref{thm:join-liveness}, every node in $J$ is either present
	at $t^*$
	or enters during
	$[t^*,t_c + D]$.
	By Lemma~\ref{lem:changing-size}(b),
	$|J| \le (1+\alpha)^3 \cdot N(t^*)$.
	
	Next we calculate a lower bound on $Q_c$.
	\begin{align*}
		|Q_c| &= \beta \cdot |{\it Members}_{p_2}^{t_c}| &\qquad \mbox{by the code} \\
		&\ge \beta \cdot [(1-\alpha)^3 - \Delta \cdot (1 + \alpha)^2]
		\cdot N(\max\{0,t_c-3D\})
		&\qquad \mbox{by Lemma~\ref{lem:members-lb}} \\
		&\ge \beta \cdot [(1-\alpha)^3 - \Delta \cdot (1 + \alpha)^2]
		\cdot (1+\alpha)^{-1} \cdot
		N(t^*)
		&\qquad \mbox{by Lemma~\ref{lem:changing-size} (b)}
	\end{align*}
	
	We now calculate a lower bound on $|K|$.
	By Lemma~\ref{lem:active-3D}, at most $(1 - Z) \cdot N(t_s)$
	nodes crash or fail during $[t_s, t_c+D]$, since the length
	of the interval is at most $3D$.
	In the worst case, all the nodes that crash or fail are in $Q_s$.
	\begin{align*}
		|K| &\ge |Q_s| - (1 - Z) \cdot N(t_s) \\
		&= \beta \cdot |{\it Members}_{p_1}^{t_s}| - (1 - Z) \cdot N(t_s) \\
		& \qquad \mbox{by the code} \\
		&\ge \beta \cdot [(1-\alpha)^3 - \Delta \cdot (1 + \alpha)^2]
		\cdot N(\max\{0,t_s - 3D\}) -
		(1 - Z) \cdot N(t_s) \\
		& \qquad \mbox{by Lemma~\ref{lem:members-lb}} \\
		&\ge \beta \cdot [(1-\alpha)^3 - \Delta \cdot (1 + \alpha)^2]
		\cdot (1 + \alpha)^{-3} \cdot N(t^*)
		- (1 - Z) \cdot N(t_s) \\
		& \qquad \mbox{by Lemma~\ref{lem:changing-size}(b) since
			$0 < t_c - t_s < 2D$} \\
		&\ge \beta \cdot [(1-\alpha)^3 - \Delta \cdot (1 + \alpha)^2]
		\cdot (1 + \alpha)^{-3} \cdot N(t^*)
		- (1 - Z) \cdot (1+\alpha)^2 \cdot
		N(t^*) \\
		& \qquad \mbox{by Lemma~\ref{lem:changing-size}(b)
			since $0 < t_s - t^* < 2D$ and $1-Z > 0$} \\
		%
		&= \left[ \beta \cdot [(1-\alpha)^3 - \Delta \cdot (1 + \alpha)^2]
		\cdot (1+\alpha)^{-3} - (1-Z) \cdot
		(1+\alpha)^2 \right] \cdot N(t^*)
	\end{align*}
	
	Finally, we show $|Q_c| + |K| > |J|$.
	\begin{align*}
		|Q_c| + |K| &\ge [\beta \cdot [(1-\alpha)^3 - \Delta \cdot (1 + \alpha)^2]
		\cdot (1+\alpha)^{-1}
		+ \beta \cdot [(1-\alpha)^3 \\
		& \hspace*{.2in} - \Delta \cdot (1 + \alpha)^2] 
		\cdot (1+\alpha)^{-3} - (1-Z) \cdot
		(1+\alpha)^2 ] \cdot
		N(t^*) \\
		&> (1 + \alpha)^3 \cdot
		N(t^*)
		\qquad \mbox{by Constraint~\ref{parameters:H}} \\
		&\ge |J|. \hfill
	\end{align*}
	\qed
\end{proof}

\begin{theorem}
\label{thm:safety}
The schedule resulting from the restriction of the execution to the
{\em store} and {\em collect} invocations and responses satisfies
regularity for the store-collect problem.
\end{theorem}

\begin{proof}
(1) Suppose $cop$ is a {\em collect} operation that returns view $V$.
Let $c$ be the collect phase of $cop$.
Let $p$ be a node.
If $V(p) = \bot$ and a {\em store} operation by $p$,
consisting of store phase $s$,
precedes $cop$, then, by Lemma~\ref{lem:views-ordered-by-time},
{\it view}$(s) \preceq$ {\it view}$(c)$.
Hence, {\it view}$(s)$ contains a tuple for $p$ with a non-$\bot$ value,
which is a contradiction.

Therefore, $V(p) = v \neq \bot$.  We show that a
{\sc Store}$_p(v)$ invocation occurs before {\it cop} completes
and no other {\em store} operation by $p$ occurs between this
invocation and the invocation of {\it cop}.
A simple induction shows that every (non-$\bot$)
value for one node in another node's {\it LView} variable
at some time comes from a
{\sc Store} invocation by the first node that has already occurred.
Since $V$ is the value of the invoking node's {\it LView} variable when
{\it cop} completes, there is a previous {\sc Store}$_p(v)$ invocation.

Now suppose for the sake of contradiction that the {\sc Store}$_p(v)$
completes---call this operation {\it sop}---and there is another
{\em store} operation by $p$, call it {\it sop}$'$,
that follows {\it sop} and precedes {\it cop}.
Let $v'$ be the value of {\it sop}$'$; by the assumption of unique
values, $v \ne v'$.
Since {\it sop} and {\it sop}$'$ are executed by the same node, it is
easy to see from the code that {\it view}$(sop) \preceq$
{\it view}$(sop')$.
By Lemma~\ref{lem:views-ordered-by-time}, {\it view}$({\it sop}')
\preceq$ {\it view}$(c) = V$.
But then value $v$ is superseded by value $v' \ne v$, contradicting
the assumption that $V(p) = v$.

(2) Suppose {\it cop}$_1$ and {\it cop}$_2$ are two
{\em collect} operations such that {\it cop}$_1$ returns $V_1$,
{\it cop}$_2$ returns $V_2$, and {\it cop}$_1$ precedes {\it cop}$_2$.
Note that {\it cop}$_1$ contains a store phase $s$ which
finishes before the collect phase $c$ of {\it cop}$_2$ begins.
By Lemma~\ref{lem:views-ordered-by-time},
{\it view}$(s) \preceq$ {\it view}$(c)$.
Regularity holds since {\it view}$(s) = V_1$ and {\it view}$(c) = V_2$,
implying that $V_1 \preceq V_2$.\qed
\end{proof}

By Theorem~\ref{thm:join-liveness}, every node that enters
and remains active sufficiently long eventually joins.
Since a {\em store} operation consists of a store phase
and a {\em collect} operation consists of a collect phase followed
by a store phase, by Theorem~\ref{thm:phase-liveness},
which states that every phase eventually completes as long as the
invoker remains active,
every operation eventually completes as long as the invoker remains active.
Finally, Theorem~\ref{thm:safety} ensures regularity.

\begin{coro}
\label{cor:correctness}
\AlgName{} is a correct implementation of a store-collect object,
in which each {\em Store} or {\em Collect} completes within a
constant number of communication rounds.
\end{coro}


\section{Implementing Distributed Objects Despite Continuous Churn}

In this section we show how to implement 
a variety of objects using store-collect.
For all applications, we assume that the conditions for store-collect
termination hold, which guarantees termination of the operations.

\subsection{Simple, Non-Linearizable Objects}
\label{section:small apps}


We start with three simple applications of store-collect for implementing
other (non-linearizable) shared objects.\footnote{
	The behavior of these objects can be formalized through
	\emph{interval linarizability}~\cite{CastanedaRR2018}.}
The choice of problems and algorithms follow \cite{KuznetsovRTP2019},
but the algorithms inherit good efficiency properties from our
store-collect implementation.

\paragraph{Max register}
\label{subsection:max register}
holds the largest value written into it~\cite{AspnesAC2012};
provides two operations:
\begin{itemize}
	\item $\mathsc{writeMax}(v)$
	takes a value $v$ as an argument and returns \textsc{Ack}.
	\item $\mathsc{readMax}()$
	has no arguments and returns a value.
\end{itemize}
Its sequential specification consists of all sequences
of \mathsc{writeMax} and \mathsc{readMax} operations
in which each \mathsc{readMax} returns the largest argument of all preceding
\mathsc{writeMax}'s, if any exists, and $0$, if there is none.

Algorithm~\ref{algorithm:max register} uses a single store-collect
object, holding a single value \textit{val} for each node,
and a local variable $V$ for each node, holding a view.
\mathsc{writeMax} stores the new value (Line \ref{line:mr1})
and returns \textsc{Ack} (Line \ref{line:mr2}).
\mathsc{readMax} collects a view (Line \ref{line:mr3}) and
returns the maximum value stored in it (Line \ref{line:mr4}).

	\begin{algorithm}[tb]
	\begin{algorithmic}[1]
		\setalglineno{55}
		\small
		\setlength{\multicolsep}{0pt}
		\begin{multicols*}{2}
			\item[] \blank{-.55cm} \bf When $\mathsc{writeMax}_p(v)$ occurs:\\
			\STATE $\mathsc{Store}_p(v)$  \label{line:mr1}
			\STATE return \mathsc{Ack}		\label{line:mr2}
			\item[]
			\item[] \blank{-.45cm}\bf When $\mathsc{readMax}_p()$ occurs:\\
			\STATE $V = \mathsc{Collect}_p()$ \label{line:mr3}
			\STATE return $\max(V.\mbox{\it val})$  \label{line:mr4}
			
		\end{multicols*}
	\end{algorithmic}
	\caption{Max-Register: code for node $p$.}
	\label{algorithm:max register}
\end{algorithm}

The main correctness properties achieved are as follows.
If any value is written before the end of the collect by a
\mathsc{readMax}, then by the regularity property of store-collect,
the \mathsc{readMax} returns the maximum value of all the values written
before it. If the start of the store by a \mathsc{writeMax} follows a
\mathsc{readMax}, then the \mathsc{readMax} does not consider the store
value.

\paragraph{Abort flag}
\label{subsection:abort flag}
a Boolean flag that can only be raised from \textit{false}
to \textit{true}~\cite{KuznetsovRTP2019};
provides two operations:
\begin{itemize}
	\item $\mathsc{abort}()$
	has no arguments and returns \textsc{Ack}.
	\item $\mathsc{check}()$
	has no arguments and returns \emph{true} or \emph{false}.
\end{itemize}
Its sequential specification consists of all sequences of
\textsc{abort} and \textsc{check} operations in which each \textsc{check}
returns \emph{true} if an \textsc{abort} precedes it, and otherwise
returns \emph{false}.

Algorithm~\ref{algorithm:abort flag} follows~\cite{KuznetsovRTP2019}.
It uses a single store-collect object,
holding a single \textit{flag} for each node,
and a local variable $F$ for each node, holding a view.
\textsc{abort} stores true (Line \ref{line:af1}) and
returns \textsc{ack} (Line \ref{line:af2}). \textsc{check}
collects the \textit{flags} (Line \ref{line:af3}).
If any of the flags is true then \textsc{check} returns true
(Line \ref{line:af4}).
Otherwise, returns false (Line \ref{line:af5}).

	
	\begin{algorithm}[H]
		\begin{algorithmic}[1]
			\setalglineno{59}
			\small
			\setlength{\multicolsep}{0pt}
			\begin{multicols*}{2}
			\item[] \blank{-.45cm}\bf When $\mathsc{abort}_p()$ occurs:\\
			\STATE $\mathsc{Store}_p(\mbox{\it true})$ \label{line:af1}
			\STATE return \mathsc{Ack}	\label{line:af2}		
			\columnbreak
			\item[] \blank{-.45cm}\bf When $\mathsc{check}_p()$ occurs:\\
			\STATE $F = \mathsc{Collect}_p()$ \label{line:af3}
			\STATE \textbf{if \mbox{\it $\exists q$ s. t. $ \textit{F(q)} == \textit{true}$} then} \label{line:af4}
			\STATE \quad return \mbox{\it true} \label{line:af5}
			\STATE \textbf{else} return \mbox{\it false} \label{line:af6}
			\end{multicols*}
		\end{algorithmic}
		\caption{Abort Flag: code for node $p$.}
		\label{algorithm:abort flag}
	\end{algorithm}
	

The main correctness property achieved is the following.
If an \textsc{abort} completes before a \textsc{check} starts,
then in particular, its store raises the flag before the
end of the collect by \textsc{check}.
Hence, the \textsc{check} returns true by the regularity property
for store-collect.
Otherwise, \mathsc{check} returns false.


\paragraph{Set}
\label{subsection:set}
contains all values added into it \cite{KuznetsovRTP2019};
provides two operations:
\begin{itemize}
	\item $\mathsc{addSet}(v)$
	takes a value $v$ as an argument and returns {\sc Ack}.
	\item $\mathsc{readSet}()$
	has no arguments and returns a set of values.
\end{itemize}
Its sequential specification consists of all sequences of \textsc{addSet}
and \textsc{readSet} operations in which \textsc{readSet} returns
the set of all values added by preceding \textsc{addSet} operations.

Algorithm~\ref{algorithm:set} uses a store-collect
object, holding a set of values for each node,
and two local variables for each node:
$S$, a view,
and \textit{LSet}, holding all values previously stored by $p$.
\textsc{addSet} adds the value to the local set (Line \ref{line:set1}),
stores it (Line \ref{line:set2}),
and returns \textsc{ack} (Line \ref{line:set3}).
\textsc{readSet} collects the set of values (Line \ref{line:set4})
and returns the union of all the sets of values (Line \ref{line:set5}).

	\begin{algorithm}[tb]
		\begin{algorithmic}[1]
			\setalglineno{65}
			\small
			\setlength{\multicolsep}{0pt}
			\begin{multicols*}{2}
			\item[] \blank{-.45cm}\bf When $\mathsc{addSet}_p(v)$ occurs:\\
			\STATE \mbox{\it LSet} $=$ \mbox{\it LSet} $\cup \{v\}$ \label{line:set1}
			\STATE $\mathsc{Store}_p$(\mbox{\it LSet}) \label{line:set2}
			\STATE return \mathsc{Ack} \label{line:set3}
			\item[]
			\item[] \blank{-.45cm}\bf When $\mathsc{readSet}_p()$ occurs:\\
			\STATE $S = \mathsc{Collect}_p()$ \label{line:set4}
			\STATE return $\cup S.$\mbox{\it set} \label{line:set5}
			\end{multicols*}
		\end{algorithmic}
		\caption{Set: code for node $p$.}
		\label{algorithm:set}
	\end{algorithm}
	
	

The main correctness property achieved is the following.
If an \textsc{addSet} stores a value $v$ before the end of a collect by
a \textsc{readSet}, then the \textsc{readSet} returns a set of values
that includes $v$ by the regularity property for store-collect.
Otherwise, the \textsc{readSet}'s return value does not include $v$.


\subsection{Atomic Snapshots}
\label{section:atomicSnap}

Like other atomic snapshot algorithms~\cite{AfekADGMG1993,DelporteFRR2018,SpiegelmanK2016},
our algorithm uses repeated collects to identify an atomic scan when two collects
return the same collected views.
Updates help scans to complete by embedding an atomic scan
that can be borrowed by overlapping scans they interfere with.
The set from which the values to be stored in the snapshot object
are taken is denoted {\it Val}$_{AS}$.
A \emph{snapshot view} is a subset of $\Pi \times {\it Val}_{AS}$,
i.e., a set of (node id, value) pairs, without duplicate node ids.

Formally, an \emph{atomic snapshot}~\cite{AfekADGMG1993}
provides two operations:
\begin{itemize}
\item
{\sc Scan}(), which has no arguments and returns a snapshot view,
and
\item
{\sc Update}($v$), which takes a value $v \in \mathit{Val}_{AS}$ as an argument and
returns {\sc Ack}.
\end{itemize}
Its sequential specification consists of all sequences of updates and scans
in which the snapshot view returned by a {\sc Scan} contains the value of
the last preceding {\sc Update} for each node $p$,
if such an {\sc Update} exists, and no value, otherwise.

An implementation should be \emph{linearizable}~\cite{HerlihyW1990}.
Roughly speaking, for every execution $\alpha$,
we should find a sequence of operations,
$\textit{lin}(\alpha)$,
containing all completed operations in $\alpha$ and some
of the pending operations,
such that:
\begin{itemize}
\item $\textit{lin}(\alpha)$
is in the sequential specification of an atomic snapshot, and
\item $\textit{lin}(\alpha)$
preserves the real-time order of
non-overlapping operations in $\alpha$.
\end{itemize}

Our algorithm to implement an atomic snapshot uses
a store-collect object, whose values are taken from the
set ($\mathcal{P}$ indicates the power set of its argument):
\begin{align*}
{\it Val}_{SC} = {\it Val}_{AS} \times \mathbb{N} \times \mathbb{N}
            \times \mathcal{P}(\Pi \times {\it Val}_{AS})
            \times \mathcal{P}(\Pi \times \mathbb{N})
\end{align*}
The first component (\textit{val}) holds
the argument of the most recent update invoked at $p$.
The second component (\textit{usqno}) holds
the number of updates performed by $p$.
The third component (\textit{ssqno}) holds
the number of scans performed by $p$.
The fourth component (\textit{sview}) holds
a snapshot view that is the result of a recent scan done by $p$;
it is used to help other nodes complete their scans.
The fifth component (\textit{scounts}) holds
a set of counts of how many scans have been done by the other nodes,
as observed by $p$.
The projection of an element $v$ in {\it Val}$_{SC}$
onto a component is denoted, respectively,
$v.\textit{val}$, $v.\textit{usqno}$, $v.\textit{ssqno}$, $v.\textit{sview}$,
$v.\textit{scounts}$.


A {\em store-collect view} is a subset of $\Pi \times {\it Val}_{SC}$,
i.e., a set of (node id, value) pairs, with no duplicate node ids.
We extend the projection notation to a store-collect view $V$,
so that $V.\textit{comp}$ is the result of replacing each tuple
$\langle p, v \rangle$ in $V$ with $\langle p, v.\textit{comp} \rangle$.
Recall that for any kind of view $V$,
$V(p)$ is the second component of the pair whose first component
is $p$ ($\bot$ if there is no such pair).
Sometimes we restrict attention to those tuples in a view $V$ whose
\textit{val} component is a ``real'' value, reflecting a store;
to this end we use the notation
\[ r(V) = \{ \langle p, v \rangle | \langle p,v \rangle \in V 
\mathrm{ and } v.\textit{val} \ne \bot\} . \]

To execute a \textsc{Scan},
Algorithm~\ref{algorithm:atomic snapshot}
increments the scan sequence number (\textit{ssqno})
(Line~\ref{line:sc1})
and stores it in the shared store-collect object
with all the other components unchanged,
indicated by the $-$ notation.
Then, a view is collected (Line~\ref{line:sc3}).
In a while loop, the last collected view is saved
and a new view is collected (Line~\ref{line:sc6}).
If the two most recently collected views reflect 
the same set of updates
(Line~\ref{line:sc7}),
the latest collected view is returned (Line~\ref{line:sc8});
Lines~\ref{line:sc7} and~\ref{line:sc8} consider only tuples with
``real'' values.
We call this a \emph{successful double collect},
and say that this is a \emph{direct scan}.
Otherwise, the algorithm checks whether the last collected view
contains a node $q$ that has observed its own \textit{ssqno},
by checking the \textit{scounts} component (Line~\ref{line:sc9}).
If this condition holds,
the snapshot view of $q$ is returned (Line~\ref{line:sc10});
we call this a \emph{borrowed scan}.

An \textsc{Update} first obtains all \text{scan sequence numbers} from
a collected view and assigns them to a local variable \textit{scounts}
(Line~\ref{line:sc11}).
Next, the value of an embedded scan is saved in a local
variable \textit{sview} (Line~\ref{line:sc12}).
Then it sets its \textit{val} variable to the argument value and increments
its update sequence number (Lines~\ref{line:sc16} and~\ref{line:sc13}).
Finally the new value, update sequence number, collected view, and set of
scan sequence numbers are stored; the node's own scan sequence
number is unchanged (Line~\ref{line:sc14}).



\begin{algorithm}[tb]
\begin{algorithmic}[1]
\setalglineno{70}
\small
\item[] {\bf Local Variables:}
\item[] \textit{ssqno}: int, initially 0
           \COMMENT{counts how many scans $p$ has invoked so far}
\item[] \textit{scounts}: set of (node id, integer) pairs with
            no duplicate node ids; initially $\emptyset$
\item[] \textit{val}: an element of {\it Val}$_{AS}$, initially $\bot$
           \COMMENT{argument of most recent update invoked by $p$}
\item[] \textit{usqno}: int, initially 0
           \COMMENT{number of updates $p$ has invoked so far}
\item[] \textit{sview}: a snapshot view, initially $\emptyset$
            \COMMENT{the result of recent embedded scan by $p$}
\item[] $V_1$, $V_2$: store-collect views, both initially $\emptyset$
\item[] \hrulefill
\setlength{\multicolsep}{0pt}
		\begin{multicols*}{2}
			\item[] \bf When {\sc scan}$_p$() occurs:\\
			\STATE $\mbox{\it ssqno}$++  \label{line:sc1}
			\STATE $\mathsc{Store}_p(\langle -, -, \mbox{\it ssqno},
                               -, - \rangle)$ \label{line:sc2}
			\STATE $V_1 = \mathsc{Collect}_p()$ \label{line:sc3}
			\WHILE{$\mbox{\it true}$}  \label{line:sc4}
			\STATE $V_2 = V_1;$  \label{line:sc5}
                $V_1 = \mathsc{Collect}_p()$  \label{line:sc6}
			\IF{($r(V_1).\mbox{\it usqno} == r(V_2).\mbox{\it usqno}$)}  \label{line:sc7}
			\STATE return $r(V_1).\mbox{\it val}$ \COMMENT{direct scan}  \label{line:sc8}
			\ENDIF
			\IF{\mbox{$\exists q$} $\mbox{\it such that}$  \\
     $\langle\mbox{\it p, ssqno}\rangle$ $\in$ $V_1(q).$\mbox{\it scounts}}
                                             \label{line:sc9}
			\STATE return $V_1(q).\mbox{\it sview}$ \\
                                      \COMMENT{borrowed scan}  \label{line:sc10}
			\ENDIF
			\ENDWHILE
			\item[]
			\item[] \bf When \mathsc{Update}$_p(v)$ occurs:\\
			\STATE $\mbox{\it scounts} = \mathsc{Collect}_p().\mbox{\it ssqno}$ \label{line:sc11}
			\STATE $\mbox{\it sview} = \mathsc{Scan}_p()$ \COMMENT{embedded scan} \label{line:sc12}
			\STATE $\mbox{\it val} = v$ \label{line:sc16}
                        \STATE \mbox{\it usqno}++   \label{line:sc13}
			\STATE $\mathsc{Store}_p(\langle \mbox{\it val, usqno, $-$, sview, scounts}\rangle)$  \label{line:sc14}
			\STATE return {\sc Ack}		\label{line:sc15}
		\end{multicols*}
	\end{algorithmic}
	\caption{Atomic snapshot: code for node $p$.}
	\label{algorithm:atomic snapshot}
\end{algorithm}

To prove linearizability, 
we consider an execution and specify an ordering
of all the completed scans and all the updates
whose store on Line~\ref{line:sc14} started.
The ordering takes into consideration the embedded scans,
which are inside updates, as well as the
``free-standing'' scans; since scans do not change the state of the atomic
snapshot object, it is permissible to do so.
%


We first show that the snapshot views returned by direct scans are
comparable in the following order:
Let $W_1$ be the snapshot view returned by a direct scan
based on the collect view $V_1$ (cf.\ Line~\ref{line:sc8})
and $W_2$ be the snapshot view returned by a direct scan
based on the collect view $V_2$ (cf.\ Line~\ref{line:sc8}).
We define $W_1 \preceq W_2$ if
for every $\langle p,v \rangle \in W_1$,
there exists $\langle p,v' \rangle \in W_2$
where the {\it usqno} associated with $v$ in $V_1$ is less than or
equal to the {\it usqno} associated with $v'$ in $V_2$.

\begin{lemma}
\label{lemma:comparable direct scans}
If a direct scan by node $p$ returns $W_1$ and
a direct scan by node $q$ returns $W_2$,
then either $W_1 \preceq W_2$ or $W_2 \preceq W_1$.
\end{lemma}

\begin{proof}
Let {\it cop}$_p^1$, returning $V_1'$,
followed by {\it cop}$_p^2$, returning $V_1$,
be the successful double collect at the end of $p$'s direct scan and
let {\it cop}$_q^1$, returning $V_2'$,
followed by {\it cop}$_q^2$, returning $V_2$,
be the successful double collect at the end of $q$'s direct scan.
Note that $W_1 = r(V_1)$.{\it val},
which is equal to $r(V'_1)$.{\it val},
and similarly $W_2 = r(V_2)$.{\it val} $= r(V'_2)$.{\it val}.

{\em Case 1:} {\it cop}$_p^1$ completes before {\it cop}$_q^2$ starts.
Consider any $\langle r,w \rangle \in W_2$.
Then $\langle r,v \rangle$ is in both $V'_1$ and $V_1$,
with $v.${\it val} $= w$ and $v.${\it usqno} $> 0$.
By regularity of store-collect, $V_1' \preceq V_2$.
Thus there is an entry $\langle r,v' \rangle \in V_2$ such that either
$v = v'$ or $v'$ is stored by $r$ after $v$ is stored by $r$.
Since the {\it usqno} variable at $r$ takes on increasing values,
$v.${\it usqno} $\le v'.${\it usqno}.
Thus there is an entry $\langle r,w' \rangle \in W_2$
where the {\it usqno} associated with $w'$ is at least as large as
that associated with $w$.
Hence $W_1 \preceq W_2$.

{\em Case 2:} {\it cop}$_q^1$ completes before {\it cop}$_p^2$ starts.
An analogous argument shows that $W_2 \preceq W_1$.
\qed
\end{proof}

Consider all direct scans in the order they complete and place them
by the comparability order.
Suppose a direct scan returning snapshot view $W_1$,
obtained from collect view $V_1$,
precedes another direct scan returning snapshot view $W_2$,
obtained from collect view $V_2$.
The regularity of store-collect ensures that $V_1 \preceq V_2$,
and thus $W_1 \preceq W_2$.
Hence, this ordering preserves the real-time order of non-overlapping
direct scans.

The next lemma helps to order borrowed scans.
%
%
%
Its statement is based on the observation that
if a scan $\textit{sop}_p$ by node $p$ borrows the snapshot view in
$V_1(q)$, then there is an update $\textit{uop}_q$ by $q$ that writes this
view (via a store).

\begin{lemma}
\label{lemma:borrowed scan}
If a scan $\textit{sop}_p$ by node $p$ borrows from a scan $\textit{sop}_q$
by node $q$, then $\textit{sop}_q$ starts after $\textit{sop}_p$ starts
and completes before $\textit{sop}_p$ completes.
\end{lemma}

\begin{proof}
Let $\textit{uop}_q$ be the update in which $\textit{sop}_q$ is embedded.
Since $\textit{sop}_p$ borrows the snapshot view of $\textit{sop}_q$,
its \textit{ssqno} appears in \textit{scounts}
of $q$'s value in the view collected in Line~\ref{line:sc6}.
The properties of store-collect imply that
the collect of $\textit{uop}_q$ (Line~\ref{line:sc11}) does not complete
before the store of $p$ (Line~\ref{line:sc2}) starts.
Hence, $\textit{sop}_q$ (called in Line~\ref{line:sc12})
starts after $\textit{sop}_p$ starts.
Furthermore, since the collect of $p$ returns the snapshot view
stored after $\textit{sop}_q$ completes (Line~\ref{line:sc14}),
$\textit{sop}_q$ completes before $\textit{sop}_p$ completes. \qed
%
%
\end{proof}


For every borrowed scan $\textit{sop}_1$, there exists a chain of
scans $\textit{sop}_2$, $\textit{sop}_3$, $\ldots$, $\textit{sop}_k$
such that $\textit{sop}_i$ borrows from $\textit{sop}_{i+1}$,
$1 \le i < k$, and $\textit{sop}_k$ is a direct scan \emph{from
which $\textit{sop}_1$ borrows}.
Consider all borrowed scans in the order they complete
and place each borrowed scan after the direct scan it borrows from,
as well as all previously linearized borrowed scans
that borrow from the same direct scan.
Applying Lemma~\ref{lemma:borrowed scan} inductively,
$\textit{sop}_k$ starts after $\textit{sop}_1$ starts
and completes before $\textit{sop}_1$ completes,
i.e., the direct scan from which a scan borrows is completely contained,
in the execution, within the borrowing scan.
This fact, together with the rule for ordering borrowed scans, implies
that the real-time order of any two scans,
at least one of which is borrowed,
is preserved since direct scans have already been shown to be ordered properly.


Finally, we consider all updates in the order their stores
(Line~\ref{line:sc14}) start.
Place each update, say {\em uop} by node $p$ with argument $v$,
immediately before the first scan whose returned view includes
$\langle p,v'\rangle$, where either $v' = v$ or $v'$ is the argument
of an update by $p$ that follows {\em uop}.
If there is no such scan, then place {\em uop} at the end of the ordering.
Note that all later scans return snapshot views that include
$\langle p,v'\rangle$, where either $v' = v$ or $v'$ is the argument
of an update by $p$ that follows {\em uop}.
This rule for placing updates ensures that the ordering satisfies the
sequential specification of atomic snapshots.

Note that if a scan completes before an update starts,
then the scan's returned view cannot include the update's value;
similarly, if an update completes before a scan starts,
then the scan's returned view must includes the update's value
or a later one.
This shows that the ordering respects the real-time order between
non-overlapping updates and scans.
The next lemma deals with non-overlapping updates.

\begin{lemma}	
\label{lemma:order updates}
Let $V$ be the snapshot view returned by a scan $\textit{sop}$.
If $V(p)$ is the value of an update $\textit{uop}_p$ by node $p$
and an update $\textit{uop}_q$ by node $q$ precedes $\textit{uop}_p$,
then $V(q)$ is the value of $\textit{uop}_q$ or a later update by $q$.
\end{lemma}

\begin{proof}
Let $\textit{sop}'$ be $\textit{sop}$ if $\textit{sop}$ is a direct
scan and otherwise the direct scan from which $\textit{sop}$ borrows.
Let $W$ be the (store-collect) view returned by the last two collects,
$\textit{cop}_1$ and $\textit{cop}_2$, of $\textit{sop}'$.

We now show that $V = W.\textit{val}$.
If $\textit{sop}' = \textit{sop}$, then $V = W.\textit{val}$
by Line~\ref{line:sc8}, since $\textit{sop}$ is a direct scan.
Otherwise, $V = W.\textit{val}$ because $W.\textit{val}$
is returned to the enclosing scan, assigned to \textit{sview},
and then stored (cf.\ Lines~\ref{line:sc12} and~\ref{line:sc14}).
This snapshot view is then propagated through the chain of borrowed-from
scans and their enclosing updates until reaching $\textit{sop}$ where
it is returned as $V$.

Since $V$ includes the value of $\textit{uop}_p$, so does $W$.
It follows that both stores
of $\textit{uop}_p$ start before $\textit{cop}_1$ completes and thus
before $\textit{cop}_2$ starts.
Since $\textit{uop}_q$ precedes $\textit{uop}_p$, the store of
$\textit{uop}_q$ at Line~\ref{line:sc14} completes before either store of
$\textit{uop}_p$ starts.
Thus the store of $\textit{uop}_q$ completes before $\textit{cop}_2$
starts, and by the store-collect property, the view $W$ returned
by $\textit{cop}_2$ must include the value of $\textit{uop}_q$
or a later update by $q$.
Since $V = W.\textit{val}$, the same is true for $V$.\qed
%
\end{proof}

Consider an update $\textit{uop}_p$, by node $p$, that follows an update
$\textit{uop}_q$, by node $q$, in the execution.
If $\textit{uop}_p$ is placed at the end of the (current) ordering because
there is no scan that observes its value or a later update by $p$, then
it is ordered after $\textit{uop}_q$.
If $\textit{uop}_p$ is placed before a scan, then the same must be true
of $\textit{uop}_q$.
By construction, the next scan after $\textit{uop}_p$ in the ordering, call it
$\textit{sop}$, returns view $V$ with $V(p)$ equal to the
value of $\textit{uop}_p$ or a later update by $p$.
By Lemma~\ref{lemma:order updates}, $V(q)$ must equal the value
of $\textit{uop}_q$ or a later update by $q$.
Thus $\textit{uop}_q$ cannot be placed after $\textit{sop}$, and thus
it is placed before $\textit{uop}_p$.

We now consider the termination property of the algorithm.
Suppose scan {\it sop}$_q$ by node $q$ contains two consecutive collects
{\it cop}$_1$, which returns $V_1$, followed by {\it cop}$_2$, which returns
$V_2$, and this double collect is unsuccessful.
Then $V_1(p).${\it usqno} is not equal to $V_2(p).${\it usqno} for some node $p$ and
$V_2$ does not contain an {\it scounts} that includes {\it sop}$_q$'s
scan sequence number {\it ssqno}.
Thus, there exists an update {\it uop}$_p$ by node $p$ that is observed by
$V_2$ but not by $V_1$.
The correctness of store-collect implies that {\it uop}$_p$ finishes after
$V_1$ starts.
Yet {\it uop}$_p$ does not include {\it sop}$_q$'s {\it ssqno}, which means
that {\it uop}$_p$ starts
before the store at the beginning of {\it sop}$_q$ completes.
Let $t$ be the time when the store at the beginning of {\it sop}$_q$ completes
and recall that $N(t)$ denotes the number of nodes present at time $t$.
Thus at most $N(t)$ updates are pending at time $t$, implying that
{\it sop}$_q$ has at most $N(t)$ unsuccessful double collects before it can
borrow a scan view.
%
Hence, {\sc Update} executes at most $O(N(t))$ collects and stores.
Putting the pieces together, we have:

\begin{theorem}
Algorithm~\ref{algorithm:atomic snapshot} is a linearizable
implementation of an atomic snapshot object.
The number of communication rounds in a {\sc Scan} or an {\sc Update}
operation is at most linear in the number of nodes present in the system
when the operation starts.
\end{theorem}

\subsection{Generalized Lattice Agreement}
\label{section:gla}


Let $\langle L, \sqsubseteq \rangle$ be a lattice,
where $L$ is the domain of lattice values, ordered by $\sqsubseteq$.
We assume a join operator, $\sqcup$, that merges lattice values.
A node $p$ calls a \mathsc{Propose} operation with a lattice input
value, and gets back a lattice output value.
The input to $p$'s $i$-th \mathsc{Propose} is denoted $v_i^p$
and the response is $w_i^p$.
The following conditions are required:
\begin{description}
\item[Validity]
Every response value $w_i^p$ is the join of some values proposed
before this response, including $v_i^p$, and all values returned to
any node before the invocation of $p$'s $i$-th \mathsc{Propose}.
\item[Consistency]
Any two values $w_i^p$ and $w_j^q$ are comparable.
\end{description}
This definition is a direct extension of \emph{one-shot} lattice
agreement~\cite{AttiyaHR1995}, following~\cite{KuznetsovRTP2019}.
The version studied in~\cite{FaleiroRRRV2012}
is weaker and lacks real-time guarantees across nodes.



Algorithm~\ref{algorithm:GLA} uses an atomic snapshot object,
in which each node stores a single lattice value (\textit{val}).
A \textsc{Propose} operation is simply an \textsc{Update}
of a lattice value which is the join of all the node's previous inputs,
followed by a \textsc{Scan} returning the analogous
values for all nodes, whose join is the output of \textsc{Propose}. 

	\begin{algorithm}[tb]
	\begin{algorithmic}[1]
		\setalglineno{85}
		\small
		\setlength{\multicolsep}{0pt}
		\begin{multicols*}{2}
			\item[] \blank{-.55cm} \bf When \mathsc{Propose}$_p(v)$ occurs:\\
			\STATE $\mbox{\it val} = \mbox{\it val} \sqcup v$\\
			\COMMENT{track previous inputs of $p$}
			\item[]
			\STATE $\mathsc{Update}_p(\mbox{\it val})$  \label{line:gla4}
			\STATE $\mbox{\it sview} = \mathsc{Scan}_p()$ \label{line:gla5}
			\STATE return $\sqcup \mbox{\it sview}.val$	\label{line:gla6}
		\end{multicols*}
	\end{algorithmic}
	\caption{Generalized lattice agreement: code for node $p$.}
	\label{algorithm:GLA}
\end{algorithm}


Validity and consistency are immediate from atomic snapshot properties.
Clearly, the algorithm terminates within $O(N)$ collects and stores,
where $N$ is the maximum number of nodes concurrently active during the
execution of \textsc{Propose}.
Since \textsc{Propose} includes one \textsc{Update} and one \textsc{Scan},
it terminates if the node does not crash or leave.

\section{Conclusion}
\label{section:concluson}

We have advocated for the usefulness of the store-collect object as a
powerful, flexible, and efficient primitive for implementing a variety of
shared objects in dynamic systems with continuous churn.
We presented a simple churn-tolerant implementation of store-collect
in which the store operation completes within one round trip and the
collect operation completes within two.
We presented an algorithm for atomic snapshots and another one
for generalized lattice agreement using atomic snapshot.
The good performance of the underlying store-collect carries over
to the latter two problems, since the values can be collected in parallel
rather than in series.
We also described some simple implementations of non-linearizable objects
(max register, abort flag, and set) using store-collect.
This assortment of applications highlights the ability to choose
whether we want to pay the price of linearizability or settle for the weaker
``regularity'' condition of store-collect.

If the level of churn is too great, our store-collect algorithm is
not guaranteed to preserve the safety property; that is, a collect might
miss the value written by a previous store, essentially by the same
counter-example as that given in~\cite{AttiyaCEKW2019}.
This behavior is in contrast to the algorithms
in~\cite{AguileraKMS2011,GilbertLS2010,KuznetsovRTP2019},
which never violate the safety property but only ensure progress
once reconfigurations cease.
In future work, we would like to either improve
our algorithm to avoid this behavior or prove that any algorithm that
tolerates ongoing churn is subject to such bad behavior.

Our correctness proof for our store-collect algorithm requires that
the parameters defining the churn rate and failure fraction
satisfy certain conditions.
These conditions imply that even in the absence of
churn the failure fraction tolerable by our algorithm
is smaller than in the static
case (namely, less than one-third versus less than one-half).
Some degradation is unavoidable when allowing for the possibility
of churn, since an argument from~\cite{AttiyaCEKW2019} can be adapted
to show that when implementing store-collect
in a system with churn rate $\alpha$, the fraction of failures
must be less than $1/(\alpha+2)$.
It would be nice to find less restrictive constraints on the parameters,
either through a better analysis or a modified algorithm, or to show
that they are necessary.

Another desirable modification to the store-collect algorithm would be
reducing the size of the messages and the amount of local storage by
garbage-collecting the {\em Changes} sets.
In the same vein, we would like to know if modifying the atomic
snapshot specification to remove from returned views entries of nodes
that have left, as is done in \cite{SpiegelmanK2016}, can lead to a
more space-efficient algorithm.

%


\section*{Acknowledgment}
We thank Luis Pantin for helpful comments.

\bibliographystyle{splncs04}
\bibliography{references}

\end{document}